\documentclass[aps,prl,10pt,a4paper,twocolumn,footinbib,superscriptaddress,showkeys]{revtex4-2}
\usepackage[utf8]{inputenc}
\usepackage{fancyhdr}
\usepackage{amsmath}
\usepackage{amssymb}
\usepackage{amsthm}
\usepackage{graphicx}
\usepackage{dsfont}
\usepackage{array, makecell}
\usepackage{yfonts}
\usepackage{subfig}
\usepackage{MnSymbol}
\usepackage[justification=Justified]{caption}
\usepackage{physics}
\usepackage{xcolor}
\usepackage{balance}
\usepackage{hyperref}

\newcommand{\trA}{\textup{tr}_{C}}

\begin{document}
\bibliographystyle{apsrev4-2}
\setcounter{secnumdepth}{2}

\title{Dissipative preparation and stabilization of \texorpdfstring{$d$}{}-mode multinomial cat states}

\author{S. Zhao}
\affiliation{Institute for Quantum Materials and Technology, Karlsruhe Institute of Technology, 76344 Eggenstein-Leopoldshafen, Germany}

\author{A. Metelmann}
\affiliation{Institute for Quantum Materials and Technology, Karlsruhe Institute of Technology, 76344 Eggenstein-Leopoldshafen, Germany}
\affiliation{Institute for Theory of Condensed Matter, Karlsruhe Institute of Technology, 76131 Karlsruhe, Germany}
\affiliation{Institut de Science et d’Ingénierie Supramoléculaires (ISIS, UMR7006), University of Strasbourg and CNRS}
 
\date{\today}
\begin{abstract}
Engineering dissipation with tailored steady states has become a powerful approach for preparing and stabilizing quantum states. In this framework, engineered dissipative processes continuously steer a system towards desired target states while suppressing unwanted noise. However, extending this idea to multimode systems is challenging and remains largely unexplored, although this class of states is a powerful resource for quantum sensing and quantum information processing applications. Here, we propose a general method to design the required dissipative processes for the generation of multimode cat states in bosonic systems. We show that the engineered dissipation prepares such states from the vacuum with high fidelity and robustly stabilizes them against decoherence. As a result, their lifetime is extended by several orders of magnitude compared to natural decay times, which in turn enhances their applications in quantum techonologies. We specifically focus on the preparation and stabilization of two-mode binomial cat states and discuss a pathway for the implementation in superconducting circuit. However, our scheme can also scale up to arbitrary $d$-mode multinomial cat states associated to $\mathfrak{su}(d\ge2)$ algebras, and thus, our scalable framework provides a feasible route towards stabilizing compact nonclassical states. 

\end{abstract}

\maketitle

\section{Introduction}
In engineered open quantum systems, dissipation can be designed such that target states span the steady-state manifold of the dynamics~\cite{PoyatosPRL96, PlenioPRL02, KrausPRL04, ValenzuelaS06, ParkinsPRL06, KrausPRA08, VerstraeteNP09, MuschikPRA11, KrauterPRL11, MirrahimiNJP14a, KapitQST17, AlbertQST19b, DoucetPRR20}, which effectively turns dissipation from a detrimental effect into a valuable resource that autonomously steers a system towards desired states while suppressing unwanted noise. Significant theoretical progress has been made for gaining a deeper insight in this direction, including mathematical properties of steady states~\cite{Albert18, NigroJSM19, KrausPRA08, FernengelJPAMT23}, relation between open-system symmetries and conserved quantities~\cite{AlbertPRA14}, and systematic reservoir design~\cite{KrausPRA08, MogilevtsevPRA13}. As a result, dissipation engineering has found broad applications in quantum information processing~\cite{KastoryanoPRL13, MarshallPRA16, HarringtonNRP22}, cooling~\cite{MurchPRL12, HacohenGourgyPRL15, MauryaPQ24}, and quantum control~\cite{MorigiPRL15, KochJPCM16, HornNJP18}.

Despite rapid theoretical progress in dissipation engineering, its experimental realization remains challenging. While single-mode bosonic implementations have been well established~\cite{RegladeN24, MarquetPRX24, LescanneNP20, BerdouPQ23, LeghtasS15, TouzardPRX18, KienzlerS15, WollmanS15}, extending these techniques to multiple bosonic modes remains difficult. Recent theoretical proposals have addressed new directions focusing on multimode cat states based on single-mode coherent state~\cite{MamaevQ18, ZapletalPQ22}, or the generalized pair coherent state \cite{AlbertQST19b}. First experimental steps towards the two-mode dissipative engineering for pair cat states~\cite{AlbertQST19b} were reported in Ref.~\cite{GertlerPQ23a}. There, the main obstacles were identified as the need for specific cross-mode nonlinear interactions and the rapidly increasing Hilbert-space complexity. In this context, realizing more complex states such as two-mode binomial cat states~\cite{AlbertPRA18} becomes rather challenging.

To illustrate this challenge associated with the dissipative preparation and stabilization of two-mode binomial cat states and other cat states within the same family, we first observe that generalized cat states constructed as superpositions of generalized coherent states are intrinsically associated to an underlying Lie algebra~\cite{PerelomovCP72, Perelomov86}. Depending on their support in the number state basis (e.g., Fock, Dicke, and spin states) or the algebra type, they are classified as non-compact (infinite support with an unbounded phase space) and compact (finite support with a bounded phase space) cat states. Some typical examples of the former is provided by the Schr\"odinger and pair cat states associated with the non-compact $\mathfrak{h}(1)$ and $\mathfrak{su}(1,1)$ algebras, whose dissipative preparation has been successfully demonstrated~\cite{MirrahimiNJP14a, AlbertQST19b, GertlerPQ23a}. In these cases, stabilization relies on the fact that these cat states are eigenstates of suitable ladder operators. In contrast, compact cat states such as the two-mode binomial cat states associated with the $\mathfrak{su}(2)$ algebra, whose number state distribution is binomial, do not admit this construction, since their finite support will be shifted rather than stabilized by ladder operators, making them incompatible with ladder-operator eigenstate conditions. Consequently, their preparation in bosonic systems typically requires other types of protocols, e.g., the unitary evolution of non-trivial initial states like Fock or NOON states~\cite{BergmannPRA16a}. Nevertheless, compact cat states can exhibit exact orthogonality between components, enabling exact quantum error correction under the Knill–Laflamme conditions~\cite{AlbertPRA18, AlbertQST19b, KnillPRL00}, whereas non-compact cat states satisfy these conditions only approximately.

In this paper, we develop a general dissipative framework for the generation and stabilization of compact generalized cat states in bosonic platforms. We demonstrate that the suitably engineered dissipation can autonomously steer the system from vacuum into the desired target cat states, while simultaneously counteracting environmental decoherence, leading to a prolonged lifetime. These stabilized cat states constitute promising resources for quantum technologies, particularly in quantum metrology~\cite{MalekiJOSABJ20b} and quantum computating~\cite{BergmannPRA16a}. As a concrete example, we focus on the dissipative preparation of two-mode binomial cat states and show that their autonomous stabilization can both extend coherent interrogating times and suppress bit-flip and phase-flip error channels. Our implementation relies on engineered nonlinear dissipation in superconducting resonators using experimentally accessible experimental elements~\cite{FrattiniAPL17b, FrattiniPRA18, ChapmanPQ23a}. Finally, we show that the construction can be generalized to arbitrary $d$-mode multinomial cat states associated with higher dimensional $\mathfrak{su}(N\geq2)$ Lie algebras, whose number state distribution follows multinomial distributions. Notably, this generalization does not explicitly require higher order nonlinear processes, thereby preserving its experimental feasibility.

\section{Two-mode binomial cat state}\label{sec_su2_cat}
\begin{figure}
        \centering
        \includegraphics[width=\linewidth]{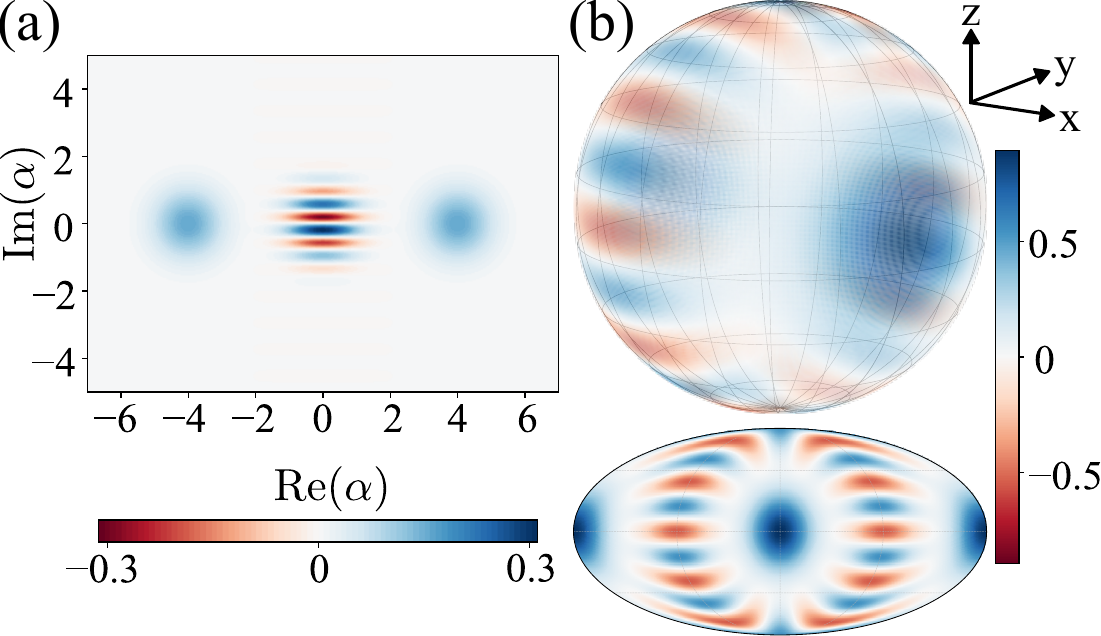}
        \caption{Wigner distribution of different types of cat states. (a) Schr\"odinger cat state $|\psi\rangle \sim |\alpha\rangle + i|-\alpha\rangle$ with $\alpha=4$. (b) Two-mode binomial cat state $|\psi\rangle \sim |N,\xi\rangle + i|N,-\xi\rangle$ with $N=15$ and $\xi=1$. Panel (a) serves as a reference for comparison, as single-mode bosonic cat states are more commonly known. Both cases exhibit the characteristic structure of two separated peaks with interference fringes in between, indicating quantum coherence between macroscopically distinct components. In (b), the state is represented on a spherical phase space, on which the second peak located near $(\theta=\pi/2,\phi=\pi)$ is less visually prominent, for which a corresponding two-dimensional projection is also shown below.}
        \label{fig_cat_state_illustration}
    \end{figure}
In this section, we briefly introduce generalized coherent states~\cite{PerelomovCP72, Perelomov86}, based on which we then define the two-mode binomial cat states. For this, we first need to understand how the shapes of phase spaces associated to different quantum systems are related to Lie algebras. Just as the infinite-dimensional quantum harmonic oscillator has a phase space represented by an unbounded complex plane (non-compact, see Fig.~\ref{fig_cat_state_illustration}(a)), finite-dimensional quantum systems possess phase spaces with bounded geometries (compact), such as a sphere (see Fig.~\ref{fig_cat_state_illustration}(b)). Representative examples with spherical phase space include atomic or spin ensembles~\cite{BonifacioPR69a, RadcliffeJPAGP71a} and a single spin-$N/2$ particle~\cite{DooleyPRA14}. Moreover, finite-dimensional systems can also be realized within intrinsically infinite-dimensional systems. For instance, two bosonic resonators coupled via a beam-splitter interaction with a fixed total excitation number (i.e., a truncated Hilbert space) effectively form a finite-dimensional system~\cite{BarzanjehPRA16a}. Here, we focus on such quantum systems whose phase space is a sphere exhibiting rotational symmetry, with Hamiltonians expressed in terms of angular momentum operators as
\begin{align}
    \hat{H} = \omega \hat{J}_z + \epsilon \hat{J}_+ + \epsilon^* \hat{J}_-, \label{eq_hamiltonian_rotation}
\end{align}
where $\omega \in \mathds{R}$ and $\epsilon \in \mathds{C}$ are system parameters. The operators $\hat{J}_z$ and $\hat{J}_\pm$ satisfy the commutation relations
\begin{align}
    [\hat{J}_+, \hat{J}_-] = 2\hat{J}_z, \quad
    [\hat{J}_z, \hat{J}_\pm] = \pm \hat{J}_\pm, \label{eq_su2_alg}
\end{align}
which define the $\mathfrak{su}(2)$ algebra. 

A generalized coherent state associated with Hamiltonians of the form Eq.~\eqref{eq_hamiltonian_rotation} is referred to as an $\mathfrak{su}(2)$ coherent state, defined as $|N,\xi\rangle = \hat{D}(\xi)|\psi_0(N)\rangle$, where $\xi \in \mathds{C}$ parametrizes the state and $\hat{D}(\xi)$ is the displacement operator, which corresponds to a rotation on the phase space sphere, and $|\psi_0(N)\rangle$ is the ground state of the system with $N \in \mathds{N}$~\cite{Klimov09}. Similarly, if the system has a non-compact phase space and its operators satisfy the $\mathfrak{su}(1,1)$ algebra, the corresponding coherent states are $\mathfrak{su}(1,1)$ coherent states, whose macroscopically distinct superpositions are known as pair cat states~\cite{AlbertQST19b, GertlerPQ23a}. In contrast, we can analogously construct the even/odd $\mathfrak{su}(2)$ cat state as \begin{align}
    |\psi_\pm\rangle=\frac{1}{\sqrt{2}}\left(|N,\xi\rangle\pm|N,-\xi\rangle\right), \label{eq_def_cat_state}
\end{align}
where $\pm$ denotes the even and odd cat states. Since a cat state requires its two constituent components $|N,\pm\xi\rangle$ to be macroscopically distinct, we first consider their overlap. By choosing $|\xi|=1$ (e.g., $\xi=1$), the two coherent states $|N,\pm 1\rangle$ become orthogonal, with
\begin{align}
    \langle N,1|N,-1\rangle = \left(\frac{1-|\xi|^2}{1+|\xi|^2}\right)^N = 0,
\end{align}
which ensures their distinguishability. The macroscopicity of the cat state is then controlled by the parameter $N \in \mathds{N}$. For sufficiently large $N$, the states become increasingly classical and well separated in phase space. In contrast, for small $N$, the states remain strongly quantum; in particular, $N=0$ gives $|0,\xi\rangle = |0\rangle \otimes |0\rangle$, corresponding to the vacuum state.

Now, we translate the notion of $\mathfrak{su}(2)$ cat states into an effective bosonic description in two resonators. We first express the $\mathfrak{su}(2)$ generators in their two-mode bosonic $N$-representation (or Schwinger representation~\cite{SakuraiCAw20a})
\begin{align}
    \hat{J}_+ = \hat{a}^\dag \hat{b}, \quad
    \hat{J}_- = \hat{a} \hat{b}^\dag, \quad
    \hat{J}_z = \frac{1}{2}(\hat{a}^\dag \hat{a} - \hat{b}^\dag \hat{b}),
    \label{eq_2_mode_rep_su2}
\end{align}
where $\hat{a}$ and $\hat{b}$ are bosonic annihilation operators acting on the two resonators. These operators satisfy the $\mathfrak{su}(2)$ commutation relations in Eq.~\eqref{eq_su2_alg}. To obtain a finite-dimensional representation, we restrict the system to a fixed total excitation number $N$ shared between the two resonators. In this subspace, the ground state can be identified as $|\psi_0(N)\rangle = |0\rangle \otimes |N\rangle$ (or equivalently $|N\rangle\otimes|0\rangle$) as desired, where $|n\rangle$ denotes a Fock state. The corresponding $\mathfrak{su}(2)$ coherent state in this bosonic realization is then given by (see appendix~\ref{app_01})
\begin{align}
    \nonumber |N,\xi\rangle &= \hat{D}(\xi) |0\rangle \otimes |N\rangle \\
    &= \frac{1}{\sqrt{(1+|\xi|^2)^N}} \sum_{n=0}^N \binom{N}{n}^{\frac{1}{2}}\xi^n \ |n\rangle \otimes |N-n\rangle,
    \label{Eq01}
\end{align}
where $\binom{N}{n}$ is the binomial coefficient, and the displacement operator $\hat{D}(\xi) = e^{\tan^{-1}(\xi)(\hat{a}^\dag \hat{b} - \hat{a} \hat{b}^\dag)}$ has the form of a beam-splitter transformation. From Eq.~\eqref{Eq01}, we see that the $\mathfrak{su}(2)$ coherent states, when realized in two-mode bosonic systems, exhibit binomial statistics in the tensored Fock state basis $|n\rangle \otimes |N-n\rangle$. We therefore refer to them as two-mode binomial coherent states. Correspondingly, the even/odd $\mathfrak{su}(2)$ cat states defined in Eq.~\eqref{eq_def_cat_state} will be referred to as two-mode binomial cat states. Similarly, coherent/cat states associated with higher-dimensional Lie algebras $\mathfrak{su}(d>2)$ exhibit multinomial statistics in their tensored Fock state basis (tensor product of $N$ Fock states), and will be referred to as $d$-mode multinomial coherent/cat states. The preparation of two-mode binomial cat states via coherent (unitary) protocols has been investigated~\cite{BarzanjehPRA16a, AlbertPRA18, MalekiEPJP21}. However, such protocols typically require nontrivial initial states, such as specific Fock or NOON states, which are themselves challenging to prepare.

In the next section, we instead propose a dissipative protocol that prepares these states from the trivial vacuum state, and stabilizes them against noise. We also demonstrate how this approach may be extended to arbitrary $\mathfrak{su}(N \ge 2)$ cat states.

\section{Preparation}\label{sec_preparation}
To motivate our dissipative preparation of two-mode binomial cat states, we first review how non-compact cat states are prepared in previous works. In general, both the Schr\"odinger cat states $|\alpha\rangle \pm |-\alpha\rangle$~\cite{MirrahimiNJP14a} and pair cat states~\cite{AlbertQST19b, GertlerPQ23a} can be stabilized using a single dissipative jump operator. This is possible because the corresponding coherent states are eigenstates of appropriate lowering operators. For instance, since $\hat{a}|\alpha\rangle = \alpha|\alpha\rangle$, its corresponding Schr\"odinger cat states $|\psi_{\mathfrak{h}(1)}\rangle \sim |\alpha\rangle + e^{i\phi}|-\alpha\rangle\sim\sum_n^\infty c_n|n\rangle$ satisfies the eigenvalue equation
\begin{align}
    \hat{a}^2 |\psi_{\mathfrak{h}(1)}\rangle = \alpha^2 |\psi_{\mathfrak{h}(1)}\rangle,
\end{align}
which leads to the jump operator $\hat{L}_{\mathfrak{h}(1)} = \hat{a}^2 - \alpha^2 \mathds{1}$~\cite{MirrahimiNJP14a}. Since $\hat{L}_{\mathfrak{h}(1)}$ annihilates only the Schr\"odinger cat states, they become the unique steady states of the open system dynamics governed by the Lindblad equation (assuming $\hat{H}=0$ for simplicity)~\cite{KrausPRA08}. Similarly, for $\mathfrak{su}(1,1)$ coherent states satisfying the eigenvalue equation $\hat{a}\hat{b}|\gamma,\delta\rangle = \gamma|\gamma,\delta\rangle$~\cite{GertlerPQ23a}, the corresponding pair cat states $|\psi_{\mathfrak{su}(1,1)}\rangle\sim |\gamma,\delta\rangle+e^{i\phi}|-\gamma,\delta\rangle\sim\sum_n^\infty c_n |n+\delta\rangle\otimes|n\rangle$ obey
\begin{align}
    \hat{a}^2 \hat{b}^2 |\psi_{\mathfrak{su}(1,1)}\rangle = \gamma^2 |\psi_{\mathfrak{su}(1,1)}\rangle,
\end{align}
yielding the jump operator $\hat{L}_{\mathfrak{su}(1,1)} = \hat{a}^2 \hat{b}^2 - \gamma^2 \mathds{1}$ \cite{AlbertQST19b}. Here, both non-compact cat states can be eigenstates of suitable ladder operators, because they involve infinite superpositions in the number state basis, and this is in contrast to two-mode binomial cat states and other compact cat states, which are finite superpositions. As a result, any ladder operator acting on the compact cat states necessarily shifts the entire superposition rather than preserving it, and therefore they cannot satisfy the eigenstate conditions of ladder operators. We emphasize that, although more general choices of jump operators are in principle possible, we restrict ourselves here to ladder operator based constructions, in order to maintain experimental feasibility. In this setting, jump operators must be non-Hermitian ladder-type operators rather than Hermitian operators that stabilize individual number states~\cite{KrausPRA08}. Consequently, compact cat states such as two-mode binomial cat states cannot, in general, be dissipatively prepared and stabilized using a single jump operator, and at least two jump operators are needed.

To construct appropriate jump operators for the two-mode binomial cat states, we now formulate the problem within a Lindblad master-equation framework. We consider a two-mode system initialized in the vacuum state $|0\rangle \otimes |0\rangle$, and seek for a Lindblad superoperator $\hat{\mathcal{L}}$ that steers the system towards the target two-mode binomial cat states. More specifically, we require the target state $\hat{\rho}_{\mathrm{tar}}$ to be the unique steady state of the dynamics, so as to exclude additional unwanted steady states. The Lindblad equation for the considered scenario can be written as
\begin{align}
    \frac{d}{dt}\hat{\rho}(t)&=\hat{\mathcal{L}}\hat{\rho}(t),
\end{align}
where
\begin{align}
    \nonumber \hat{\mathcal{L}}\hat{\rho}(t)&=-i[\hat{H},\hat{\rho}(t)]+\sum_{j}\gamma_j \bigg[\hat{L}_j\hat{\rho}(t)\hat{L}_j^\dag\\
    &\hspace{1cm}-\frac{1}{2}\left(\hat{L}_j^\dag\hat{L}_j\hat{\rho}(t)+\hat{\rho}(t)\hat{L}_j^\dag\hat{L}_j\right)\bigg],
\end{align}
is the Lindbladian superoperator, $\hat{H}$ is the system Hamiltonian, and $\hat{L}_j$ is the $j$-th jump operator with the jumping rate $\gamma_j$.

At this stage, realizing $\hat{\rho}(t\rightarrow\infty)=\hat{\rho}_{\mathrm{tar}}=|\psi_\pm\rangle\langle\psi_\pm|$ may appear challenging. However, the problem can be significantly simplified by a few observations. First, since we focus only on dissipative engineering, we assume a simple system Hamiltonian $\hat{H}=\omega_a\hat{a}^\dag\hat{a}+\omega_b\hat{b}^\dag\hat{b}$. By moving to a rotating frame, we can set $\hat{H}=0$ without loss of generality. Second, since the target state is pure, $\hat{\rho}_{\mathrm{tar}}=|\psi_\pm\rangle\langle\psi_\pm|$, the steady state condition reduces to finding jump operators $\hat{L}_j$ that annihilate $|\psi_\mathrm{tar}\rangle$, i.e., $\hat{L}_j|\psi_{\mathrm{tar}}\rangle=0$. This follows from Theorem (2) in~\cite{KrausPRA08}, which equivalently states that for jump operators made from ladder operators, the steady states are pure states that can be simultaneously annihilated by all jump operators (if they exists). With these simplifications, the task reduces to constructing appropriate annihilation conditions. A suitable choice is
\begin{align}
    \nonumber \hat{L}_{1,+}^{(2)}&=\sqrt{\kappa_1}\hat{a}^\dag(\hat{a}^\dag\hat{a}+\hat{b}^\dag\hat{b}-N),\\
    \hat{L}_2^{(2)}&=\sqrt{\kappa_2}(\hat{a}^2\pm\hat{b}^2), \label{Eq02}
\end{align}
where $\hat{L}_2^{(2)}=\hat{a}^2+\hat{b}^2$ corresponds to the case of purely imaginary amplitude $\xi$ (i.e., $\xi=i$). We emphasize that this choice is not unique, and is primarily motivated by experimental simplicity and feasibility. In the following, we explain the conceptual procedure to construct these jump operators.

We first focus on the jump operator $\hat{L}_{1,+}^{(2)}$, whose role is to stabilize the system within the subspace of fixed total excitation number required for the two-mode binomial cat states. This fixed total excitation number $N$ is reflected by the fact that the operator $\hat{J}_c = \hat{a}^\dag \hat{a} + \hat{b}^\dag \hat{b}$ commutes with all generators in Eq.~\eqref{eq_2_mode_rep_su2}, and hence acts as the Casimir element of the $\mathfrak{su}(2)$ algebra. By Schur’s lemma~\cite{Serre77}, it acts trivially within the target manifold, \begin{align}
    \hat{J}_c |\psi_\pm\rangle = N |\psi_\pm\rangle.
\end{align}
This suggests that we may use $(\hat{J}_c - N)$ as a syndrome operator to detect deviations from the desired total excitation number. However, since $\hat{J}_c$ is Hermitian, it does not induce any dissipative dynamics by itself. To actively steer the system towards the correct subspace, it is combined with a creation process, leading to the form $\hat{L}_{1,+}^{(2)} \sim \hat{a}^\dag(\hat{J}_c - N)$. Physically, $(\hat{J}_c - N)$ checks whether the system has the correct number of excitations, while $\hat{a}^\dag$ injects photons when this condition is not satisfied. In this sense, $\hat{L}_{1,+}^{(2)}$ acts as a feedback mechanism that pumps the system towards the target excitation sector. In addition, this mechanism alone stabilizes the system only when the initial total excitation number is below $N$, driving it towards the desired subspace with that of $N$. However, it does not remove excess excitations if the system populates states in subspaces of larger excitations than $N$ (e.g., due to thermal noise). In that situation, the same process continues to inject excitations and therefore drives the system further away from the target subspace. To ensure stability from both sides, one can introduce a complementary loss-type jump operator $\hat{L}_{1,-}^{(2)} \sim \hat{a}(\hat{J}_c - N)$, which removes excess photons when the excitation number exceeds $N$. We would also like to highlight, that in the equivalent single-mode picture, i.e., $\hat{L}_{1,+}^{(1)} \sim \hat{a}^\dag(\hat{a}^\dag \hat{a} - N)$ and $\hat{L}_{1,-}^{(1)} \sim \hat{a}(\hat{a}^\dag \hat{a} - N)$ can be used to dissipatively prepare and stabilize the system to Fock states. Moreover, their combined jump operator $\hat{L}_1^{(1)}\sim(\hat{a}^\dag+\hat{a})(\hat{a}^\dag \hat{a}-N)$ comes with the additional feature that it can steer any arbitrary initial state into the corresponding fock state $|N\rangle$.

Then, we explain the intuition behind the second jump operator $\hat{L}_2^{(2)}$, whose role is to fix the relative structure between the two modes within the constrained excitation manifold. To this end, we examine the action of the annihilation operators on the two-mode binomial coherent states \begin{align}
    \nonumber \hat{a}|N,\pm\xi\rangle &= (\pm \xi)\sqrt{\frac{N}{2}} |N-1,\pm\xi\rangle,\\
    \hat{b}|N,\pm\xi\rangle &= \sqrt{\frac{N}{2}} |N-1,\pm\xi\rangle.
\end{align}
Both operators reduce the excitation number in the same way, but differ by a relative phase between the two branches. This suggests that we may construct linear combinations of $\hat{a}$ and $\hat{b}$ to capture this structure. In particular, \begin{align}
    \hat{L}_{2,\pm\xi}^{(2)} \sim \frac{1}{\pm\xi}\hat{a} - \hat{b}
\end{align}
annihilates the corresponding coherent state $|N,\pm\xi\rangle$. It is also worth noting that together with $\hat{L}_{1,+}^{(2)}$, this jump operator can steers and stabilize the system towards a two-mode binomial coherent state, serving as a dissipative alternative to the coherent beam-splitter preparation scheme~\cite{BarzanjehPRA16a, AlbertPRA18, MalekiEPJP21}. However, for the two-mode binomial cat states that are superpositions of $|N,\pm\xi\rangle$ (see Eq.\eqref{eq_def_cat_state}), the linear order combinations of $\hat{a}$ and $\hat{b}$ are insufficient, for they distinguish between the two components, and will drive the system toward only one branch. To stabilize the superposition, the jump operator must not be sensitive to this sign structure. This can be achieved using their quadratic order combinations, which eliminate the phase dependence. For example, when $\xi=1$, one finds
\begin{align}
    \hat{a}^2 |N,\pm1\rangle = \hat{b}^2 |N,\pm1\rangle = \frac{\sqrt{N(N-1)}}{2} |N-2,\pm1\rangle,
\end{align}
which leads to the condition
\begin{align}
    (\hat{a}^2 - \hat{b}^2) |N,\pm1\rangle = 0.
\end{align}
This directly motivates the choice of the second jump operator $\hat{L}_2^{(2)} \sim \hat{a}^2 - \hat{b}^2$. In addition, we stress that the above construction principles used for designing $\hat{L}_{1,2}$ can be extended to $d$-mode multinomial cat states associated with higher dimensional Lie algebras $\mathfrak{su}(d>2)$ as discussed in section~\ref{sec_generalization}.

\begin{figure}
    \centering
    \includegraphics[width=\linewidth]{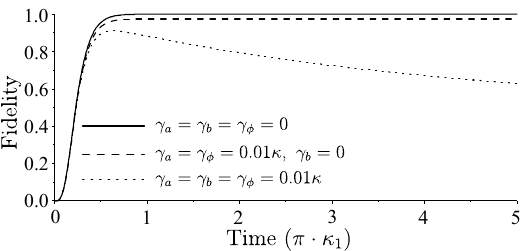}
    \caption{Evolution from the initial vacuum state $|0\rangle\otimes|0\rangle$ to the even two-mode binomial cat state $|\psi_+\rangle$ under the engineered dissipation in Eq.~\eqref{Eq02}, with $\hat{H}=0$ and $N=8$. While here we have set $\kappa_1=\kappa_2=\kappa=1$ for simplicity, in general, the convergence timescale is $\sim 1/\min\{\kappa_1,\kappa_2\}$. The parameters $\gamma_a$ and $\gamma_b$ denote single-photon loss rates in modes $a$ and $b$, respectively, and $\gamma_\phi$ denotes the (assumed equal) dephasing rate for both modes. All rates are given in units of $\kappa$.}
    \label{Fig_02}
\end{figure}
To illustrate the dissipative dynamics generated by $\hat{L}_{1,2}$ in Eq.~\eqref{Eq02}, we numerically simulate the open system evolution starting from the vacuum state, and compute the fidelity with respect to the target two-mode binomial cat state. The results are shown in Fig.~\ref{Fig_02}, where we consider the even two-mode binomial cat state $|\psi_+\rangle$. In the absence of additional noise, the system is driven into the target state with high fidelity. When including unwanted loss and dephasing processes described by the jump operators $\sqrt{\gamma_a}\hat{a}$, $\sqrt{\gamma_\phi}\hat{a}^\dag\hat{a}$, and $\sqrt{\gamma_\phi}\hat{b}^\dag\hat{b}$, the steady state becomes mixed and the fidelity saturates below unity. This can be understood from the fact that the null spaces of these noise operators and the engineered jump operators in Eq.~\eqref{Eq02} do not intersect, so that the stationary state has a little bit of impurity and contribution from outside the subspace spanned by two-mode binomial cat states~\cite{KrausPRA08}. Moreover, the dynamical behavior further reflects the interplay between engineered dissipation and noise when including the single-photon loss $\sqrt{\gamma_b}\hat{b}$. With this extra noise channel, phase-flip errors are induced within the cat-state subspace that cannot be corrected by the engineered dissipation. Thus, the fidelity exhibits a characteristic behavior in time-continuous quantum error correction: an initial rapid rise on a timescale $\sim 1/\min\{\kappa_1,\kappa_2\}$, followed by a slow decay on a timescale $\sim 1/\gamma_b$ (see Appendix~\ref{app_04} and Refs.~\cite{IppolitiPRA15, ReiterNC17}). This is in contrast to the single-photon loss $\sqrt{\gamma_a}\hat{a}$ that does not induce the phase-flip errors, and thus can be effectively corrected by the engineered dissipation.

Finally, we note that the parity of the stabilized cat state depends on the total excitation number $N$. Starting from the vacuum, the dissipative dynamics in fact may prepare either even or odd two-mode binomial cat states, whose parity depends on $N$
\begin{align}
    |\psi_\pm\rangle \sim (+1)^N|N,1\rangle + (-1)^N |N,-1\rangle.
\end{align}
This parity dependence arises from the repeated action of the creation process in $\hat{L}_{1,+}^{(2)}$, which introduces a relative phase between the two coherent components, i.e.,
\begin{align}
    \hat{a}^\dag|N,\pm1\rangle=(\pm1)\sqrt{\frac{2}{N+1}}\hat{a}^\dag\hat{a}|N+1,\pm1\rangle. \label{eq_su2_final_parity}
\end{align}
Note that although the action of the $\hat{a}^\dag$ brings us to a different state in the manifold with total excitation number $N+1$, the repeated action of $\hat{L}_{1,+}^{(2)}$ together with the combination of $\hat{L}_2^{(2)}$ will further reshape the state to the desired two-mode binomial cat states within the manifold. From a quantum error-correction perspective, this corresponds to a controlled phase flip between the even and odd cat subspaces, which can help counteract another phase-flip induced by unwanted noise channels such as single-photon gain and loss (see Sec.~\ref{sec_quan_computing}). Alternatively, this parity dependence of the final two-mode binomial cat states can be understood as the preservation of the excitation number parity in the second mode, that is, the parity operator $(-1)^{\hat{b}^\dag \hat{b}}$ commutes with other engineered jump operators. For example, if $N=\text{even}$, then the parity of $a$-mode also becomes even using Eq.~\eqref{eq_su2_final_parity}, which ensures that the even parity (the initial state is vacuum) in the $b$-mode is preserved, i.e., the Fock state in $b$-mode is $N-\text{even}=\text{even}$. This parity preservation property can be generalized to arbitrary mode multinomial cat states as shown later in table~\ref{table_summary_jump_operators}.

\subsection{Influence of loss and cat size}
In the following, we provide additional discussion on the $N$-dependence (size of the cat state) of the dissipative dynamics. From Eq.~\eqref{Eq01}, the action of the annihilation operators $\hat{a}$ and $\hat{b}$ on the two-mode binomial coherent states consists of two contributions: (i) transitions between adjacent manifolds, $|N,\pm\xi\rangle \rightarrow |N-1,\pm\xi\rangle$, and (ii) an accompanying normalization-induced prefactor scaling as $\sim \sqrt{N}$. The latter effectively enhances the operator weight over the relevant Hilbert space, similarly to how the weight of annihilation operator over single-mode coherent states $|\alpha\rangle$ scales as $\hat{a}\sim |\alpha|$, i.e., $\hat{a}|\alpha\rangle=\alpha|\alpha\rangle$. As a simple example, consider the single-photon loss channel $\sqrt{\gamma_a}\hat{a}$. Its effective strength can be quantified via its truncated operator norm over the coding subspace~\cite{Winter17, ShirokovMs20, Becker25} \begin{align}
    ||\sqrt{\gamma_a}\hat{a}|| = \max_{\psi}\sqrt{\gamma_a \langle \psi|\hat{a}^\dagger \hat{a}|\psi\rangle} \sim \sqrt{\gamma_a N}.
\end{align}
This shows that increasing $N$ effectively plays a role analogous to increasing the loss rate $\gamma_a$, in the sense that both lead to the same scaling of the effective operator strength (or effective decay rate) $\gamma_\mathrm{eff}^{(a)}\sim N\gamma_a$. More generally, composite dissipative processes with jump operators of the form $\hat{a}^p \hat{a}^{\dag q}$ scale as $\sim N^{(p+q)/2}$. This means that $\|\hat{L}_{1,+}^{(2)}\|\sim N^{3/2}$, $\|\hat{b}\| \sim \sqrt{N}$, and $\{\|\hat{a}^\dagger\hat{a}\|,\, \|\hat{b}^\dagger\hat{b}\|,\, \|\hat{L}_2^{(2)}\|\}\sim N$.

\begin{figure}
    \centering
    \includegraphics[width=\linewidth]{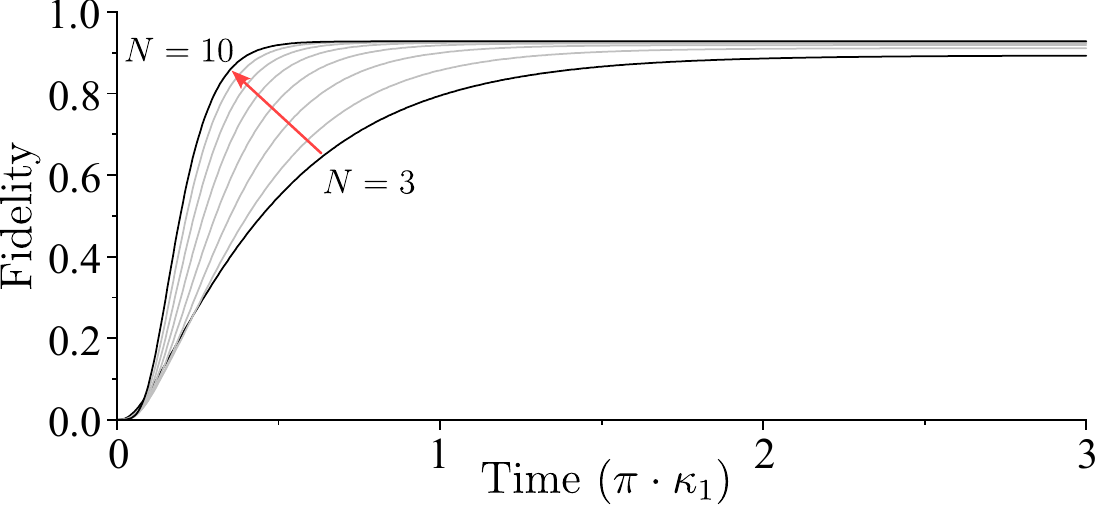}
    \caption{Fidelity evolutions of a system initialized in the vacuum state, evolving under the engineered dissipations (with rates $\kappa_{1,2}$) and unwanted noise $\{\sqrt{\gamma_a}\hat{a}, \ \sqrt{\gamma_\phi}\hat{a}^\dag\hat{a}\}, \ \sqrt{\gamma_\phi}\hat{b}^\dag\hat{b}\}$. We have set $\hat{H}=0$, $\kappa_1=\kappa_2=\kappa$, and $\gamma_a=\gamma_\phi=0.03\kappa$. Each curve corresponds to a target two-mode binomial cat state of different $N$, which runs from $N=3$ (bottom curve) to $N=10$ (top curve) with an increment of $\Delta N=1$ (indicated by the red arrow).}
    \label{Fig_03}
\end{figure}

While increasing $N$ enhances the effective strength of intrinsic noise channels such as single-photon loss and dephasing, it simultaneously amplifies the engineered dissipation channels. Importantly, the effective strength of $\hat{L}_{1,+}^{(2)}$ grows faster than all considered noise processes, while $\hat{L}_2^{(2)}$ scales comparably to dephasing. That is, in the large-$N$ regime the engineered dissipation dominates over unwanted noise, leading to improved stabilization rather than degradation of the steady state. This observation highlights that larger values of $N$ are not only desirable from the perspective of generating macroscopically distinguishable superpositions, but can also be beneficial for stabilization using engineered dissipation. In addition, such scaling behavior also emphasizes an inherent trade-off: higher-order engineered dissipation generally exhibits stronger energy-dependent scaling, but are also increasingly challenging to realize experimentally due to their rapidly decreasing contributions in superconducting implementations. Fig.~\ref{Fig_03} shows how an increasing $N$ strengthens the engineered dissipation more quicker, reducing the preparation time of the target two-mode binomial cat states.

However, as shown in the same figure, the stationary fidelity converges to a value below unity with increasing $N$. While $\hat{L}_{1,+}^{(2)}$ initially steers the system efficiently toward the target manifold with fixed total excitation number $N$ at a rate scaling as $\sim N^{3/2}$, its truncated operator norm near this manifold exhibits a weaker scaling behavior, i.e., $\|\hat{L}_{1,+}^{(2)}\|\sim N^{1/2}$. As a result, the effectiveness of the dissipative stabilization decreases in the vicinity of the target manifold, leading to a stationary fidelity that saturates below unity in the large $N$ limit. This can be alternatively understood from the fact that the syndrome operator $\hat{a}^\dag \hat{a}+\hat{b}^\dag\hat{b}-N$ scales as $\sim 1$ around the stabilized manifold, which in turn modifies the effective contribution of $\hat{L}_{1,+}^{(2)}$ in the truncated space, leading to the scaling $\|\hat{L}_{1,+}^{(2)}\|\sim\hat{a}^\dagger \sim N^{1/2}$. This reduction in scaling behavior further restricts the performance of $\hat{L}_2^{(2)}$, since the target state stabilization relies on the interplay between the two processes, making both the engineered dissipation and the unwanted noise exhibit the same scaling behavior around the target state manifold. Consequently, the fidelity of the prepared target states saturates in the limit $N \to \infty$, as shown in Fig.~\ref{Fig_03}.

\subsection{Circuit design}
\begin{figure}
    \centering
    \includegraphics[width=\linewidth]{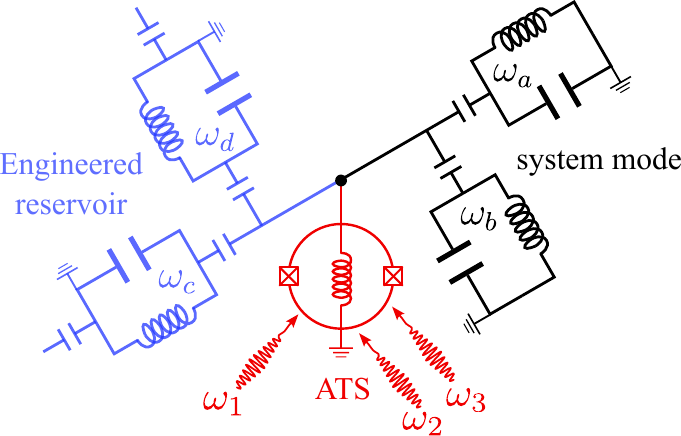}
    \caption{Illustration of four superconducting LC resonators coupled via an ATS nonlinear element. The two resonators with frequencies $\omega_a$ and $\omega_b$ form the system where the two-mode binomial cat states are encoded. The other two resonators (blue) with frequencies $\omega_c$ and $\omega_d$ act as engineered environments: they are strongly coupled to external ports (with rates $\kappa_c$ and $\kappa_d$), allowing excitations to quickly leak out so that these modes effectively behave like vacuum reservoirs. The ATS mediates nonlinear interactions between all four resonators. By driving the lossy modes with three tones at frequencies $\omega_\mathrm{dr}^{(1,2,3)}$, these interactions can be selectively activated to realize the desired dissipative processes (see Eq.~\eqref{Eq03}).}
    \label{Fig_01}
\end{figure}

Having established the form of the engineered dissipation in Eq.~\eqref{Eq02}, we now turn to its physical implementation. We consider two resonators with frequencies $\omega_a$ and $\omega_b$ to host the target two-mode binomial cat states, while two other resonators with frequencies $\omega_c$ and $\omega_d$ to act as the engineered dissipative modes (see Fig.~\ref{Fig_01}).
The desired dissipative processes in Eq.~\eqref{Eq02} can be implemented by coherently coupling the system modes to the two lossy auxiliary modes. This leads to an effective Hamiltonian of the form $\hat{H}=\hat{H}_1+\hat{H}_2$, where \begin{align}
    \nonumber \hat{H}_1&=g_1\hat{L}_{1,+}^{(2)}\hat{c}^\dag+h.c.=g_1\hat{a}^\dag(\hat{a}^\dag\hat{a}+\hat{b}^\dag\hat{b}-N)\hat{c}^\dag+h.c.,\\
    \hat{H}_2&=g_2\hat{L}_2^{(2)}\hat{d}^\dag+h.c.=g_2(\hat{a}^2\pm\hat{b}^2)\hat{d}^\dag+h.c.. \label{Eq03}
\end{align}
Physically, the terms involving $\hat{c}^\dag$ and $\hat{d}^\dag$ describe processes where excitations are transferred from the system into the lossy modes, which then quickly dissipate them into the environment. In other words, the auxiliary modes are assumed to remain close to their vacuum state, processes in the Hermitian-conjugate $h.c.$ terms involving $\hat{c}$ and $\hat{d}$ are effectively suppressed, as the excitations would already be lost into the environment before they could be transferred back. In this regime, the auxiliary resonators act as “drains” that continuously absorb excitations from the system and irreversibly dissipate them into the environment, thereby enforcing the desired dissipative structure in the system modes. Eventually, the dynamics induced by Hamiltonian Eq.~\eqref{Eq03} coincides with that of Lindblad jump operators in Eq.~\eqref{Eq02}, with the effective decay rates $\kappa_1=\frac{4g_1^2}{\kappa_c}$ and $\kappa_2=\frac{4g_2^2}{\kappa_d}$ (see appendix~\ref{app_02}).

Examining the structure of the target Hamiltonians Eq.~\eqref{Eq03}, we find that $\hat{H}_1$ consists of two- and four-wave mixing processes, whereas $\hat{H}_2$ contains three-wave mixing terms. To realize these wave-mixing processes, we consider a system consisting of four superconducting resonators coupled via an asymmetric-threaded SQUID (ATS)~\cite{LescanneNP20},  as depicted in Fig.~\ref{Fig_01}. The ATS provides a tunable nonlinear potential of the sinusoidal form, enabling a wide range of wave-mixing processes. In our scheme, we utilize the third and fourth order nonlinearities. We emphasize that this is not the only possible circuit realization, similar interactions could in principle also be implemented using other devices, such as the Josephson ring modulator~\cite{BergealNP10, BergealN10}, or a SNAIL \cite{FrattiniPRA18}.
 
However, we focus on the ATS device to generate the required interactions. Crucially, some additional constraints must also be satisfied. In particular, one must account for the freedom in choosing the sign $\pm$ in $\hat{H}_2$, as well as suppress unwanted parametric processes, such as self- and cross-Kerr terms arising naturally from the fourth-order nonlinearity which generally cannot be eliminated within the rotating-wave approximation (RWA). To address these issues, we introduce additional pumpings on the ATS device. With an appropriate choice of resonator and drive frequencies, the application of the RWA reduces the full circuit Hamiltonian to the desired effective form in Eq.~\eqref{Eq03}. A detailed derivation is provided in Appendix~\ref{app_03}. In summary, we first tune the external fluxes threading the ATS device such that the resonant Kerr-type interactions would always acquire some additional driving frequencies. This allows these unwanted terms to be eliminated within the rotating-wave approximation (RWA). We then choose the drive frequencies shown in Fig.~\ref{Fig_01} to satisfy \begin{align}
    \nonumber \omega_1 &= \omega_a + \omega_c,\\
    \nonumber \omega_2 + \omega_1 &= \omega_d - 2\omega_a,\\
    \omega_3 + \omega_1 &= \omega_d - 2\omega_b,
    \label{eq_frequency_matching}
\end{align}
thereby rendering all parametric processes appearing in Eq.~\eqref{Eq03} resonant in the interaction frame with respect to the free system Hamiltonian (i.e., the terms proportional to $\hat{a}^\dagger\hat{a}$, $\hat{b}^\dagger\hat{b}$, etc.). The driving frequency $\omega_{1}$ in Eq.~\eqref{eq_frequency_matching} realizes the parametric processes associated with $\hat{H}_1$, i.e., effectively a two-mode squeezing interaction between the auxiliary mode $c$ and mode $a$. To realize the two distinct three-photon parametric processes associated with $\hat{H}_2$, namely $\hat{a}^2\hat{d}^\dag+h.c.$ and $\hat{b}^2\hat{d}^\dag+h.c.$, two additional driving pumps with frequencies $\omega_2$ and $\omega_3$ are introduced in the second and third lines of Eq.~\eqref{eq_frequency_matching}. However, with the configuration designed to remove unwanted Kerr-type interactions, the effective resonance conditions are achieved by applying driving frequencies shifted by $\omega_1$, i.e., $\omega_2+\omega_1$ and $\omega_3+\omega_1$. We find that these combined pump frequencies can activate only the desired three-photon parametric processes without inducing any unwanted terms. Finally, by appropriately tuning the remaining system parameters, we may independently control the integer parameter $N$ appearing in $\hat{H}_1$, as well as select the desired sign, $\pm$, in $\hat{H}_2$. Moreover, although the realization of $\hat{H}_1$ (first line of Eq.~\eqref{eq_frequency_matching}) induces some additional resonant terms, we find that they do not influence the effective system dynamics (see appendix~\ref{app_03}).

\subsection{Generalization to higher dimensional Lie algebras}\label{sec_generalization}
The construction principle used in this work for two-mode binomial cat states can be directly extended to $d$-mode multinomial cat states associated to the $\mathfrak{su}(d>2)$ Lie algebras. The corresponding $d$-mode multinomial coherent states can be written as $|N,\xi_1,\dots,\xi_{d-1}\rangle$, involving $d$ bosonic modes with their annihilation operators denoted as $\hat{a}_{1,2,\dots,d}$. Here, $N$ represents the total excitation number within the system, whose corresponding observable operator is denoted by the Casimir operator $\hat{J}_c^{(d)}$ of the $\mathfrak{su}(d)$ algebra~\cite{MathurJMP01}.
A key property of these states is that all annihilation operators act in a unified manner,
\begin{align}
    \hat{a}_j |N,\xi_1,\dots,\xi_{d-1}\rangle = c_j f(N) |N-1,\xi_1,\dots,\xi_{d-1}\rangle, \label{eq_su_d_annihilation_action}
\end{align}
where $c_{j<d}=\xi_j$and $c_d=1$, and the function $f(N)$ is common to all modes. This shared structure allows us to generalize the dissipative construction in a systematic way. In particular, we can extend the stabilization operator of total excitation number in Eq.~\eqref{Eq02} to the $d$-mode as \begin{align}
    \hat{L}_{1,+}^{(d)}&=\hat{a}^\dag\left(\sum_{j=1}^d \hat{a}_j^\dag\hat{a}_j-N\right)=\hat{a}^\dag\left(\hat{J}_c^{(d)}-N\right),
\end{align}
which enforces the effective confinement of the system to the fixed total excitation subspace.

However, fixing the total excitation number alone is not sufficient to uniquely select the desired binomial statistics within this manifold. One must further stabilize the relative actions between annihilation operators, i.e. the actions of $\hat{a}_j$ and $\hat{a}_k$ differ by a factor of $\frac{c_j}{c_k}$, which requires an additional set of $d-1$ jump operators $\hat{L}_{j\geq2}^{(d)}$. More specifically, these operators are originated to some generalized eigenvalue problems. Using Eq.~\eqref{eq_su_d_annihilation_action}, we first express the action of $\hat{a}_j$ and $\hat{a}_{j-1}$ on one of the cat state component $|N,\xi_1,\dots,\xi_{d-1}\rangle$ as \begin{align}
    \begin{cases}
    \hat{a}_j |N,\xi_1,\dots,\xi_{d-1}\rangle = c_j f(N), |N-1,\xi_1,\dots,\xi_{d-1}\rangle,\\
    \hat{a}_{j-1} |N,\xi_1,\dots,\xi_{d-1}\rangle = c_{j-1} f(N), |N-1,\xi_1,\dots,\xi_{d-1}\rangle.
    \end{cases}
\end{align}
Here, we note that the other cat state component $|N,\xi_1',\dots,\xi_{d-1}'\rangle$ can have some or all of the amplitudes $\xi_\bullet'$ flipped to $-\xi_\bullet$, and for simplicy, we focus on the case $\xi_j'=-\xi_j$ for all $j=1,\dots,d-1$. With this, we avoid the need to square the annihilation operators to eliminate the $\pm$ ambiguity associated with cat states composed of superpositions of $\pm \xi_\bullet$ states, i.e., $c_jc_{j-1}=(\pm\xi_j)(\pm\xi_{j-1})=|\xi_j\xi_{j-1}|$ for $j=1,\dots,d-1$. Otherwise, for the general case, we can equivalently formulate the above relations in terms of $\hat{a}_j^2$ and $\hat{a}_{j-1}^2$. Now, multiplying the first equation by $c_{j-1}$ and the second by $c_j$ yields \begin{align}
    \nonumber c_{j-1}\hat{a}_j |N,\xi_1,\dots,\xi_{d-1}\rangle &= c_j c_{j-1} f(N), |N-1,\xi_1,\dots,\xi_{d-1}\rangle,\\
    &= c_j \hat{a}_{j-1} |N,\xi_1,\dots,\xi_{d-1}\rangle,
\end{align}
which can be rewritten as \begin{align}
    (c_{j-1}\hat{a}_j - c_j \hat{a}_{j-1}) |N,\xi_1,\dots,\xi_{d-1}\rangle = 0.
\end{align}
Using this, we can identify the jump operator \begin{align}
    \hat{L}_{j}^{(d)} = c_{j-1}\hat{a}_j - c_j \hat{a}_{j-1},
\end{align}
which annihilates the corresponding cat-state manifold. Repeating this process for all adjacent mode pairs, we obtain \begin{align}
    \hat{L}_{j=2,\dots,d}^{(d)}&=\begin{cases}
      c_{j-1}\hat{a}_{j}-c_{j}\hat{a}_{j-1} \quad 2\leq j\leq d-1,\\
      c_{j-1}^2\hat{a}_{j}^2-c_{j}^2\hat{a}_{j-1}^2 \quad j=d,
    \end{cases}
\end{align}
where the quadratic jump operator for the $j=d$ case arises from eliminating the $\pm$ dependence in the last adjacent mode pair $c_{d-1}c_d=\pm\xi_{d-1}$, by squaring their corresponding annihilation operators $\hat{a}_{d-1}^2$ and $\hat{a}_d^2$. The function of $\hat{L}_{j=2,\dots,d}^{(d)}$ is to progressively lock the relative coherence between the adjacent modes when being acted by their corresponding annihilation operators. We note that since the relative structures among the $d$ bosonic modes is transitive, it is sufficient to enforce the constraints between adjacent modes. The consistency of the structure for non-adjacent modes then follows automatically by transitivity. Physically, these jump operators can be interpreted in close analogy to $\hat{L}_{1,+}^{(2)}$ and $\hat{L}_2^{(2)}$ introduced earlier in Eq.~\eqref{Eq02}. That is, the jump operator $\hat{L}_{1,+}^{(d)}$ enforces a restriction of the steady state dynamics to the target manifold with fixed total excitation number $N$. Whereas the remaining jump operators $\hat{L}_{j\geq2}^{(d)}$ constrain the relative structures between adjacent modes within this fixed excitation manifold, thereby selecting the desired coherence (or off-diagonal entries of the density matrices) of the steady state subspace.

As a short overview for the several Lindblad jump operators discussed in this section, as well as their corresponding initial and steady states, we refer to Table~\ref{table_summary_jump_operators}. 
\begin{table*}[t]
\renewcommand{\arraystretch}{1.5}
\renewcommand{\cellgape}{\Gape[0.1cm]}
\begin{tabular}{|>{\centering\arraybackslash}m{0.03\linewidth}|>{\centering\arraybackslash}m{0.42\linewidth}|>{\centering\arraybackslash}m{0.22\linewidth}|>{\centering\arraybackslash}m{0.28\linewidth}|}
\hline
 & Jump operators & General initial state & Steady states \\
\hline
(1)
&
$
\hat{L}_1^{(1)}\sim(\hat{a}^\dag+\hat{a})(\hat{a}^\dag\hat{a}-N)$
&
$\sum_{n=0}^{\infty} f_n|n\rangle$
&
$|N\rangle$ \\
\hline

(2)
&
$\hat{L}_{1,+}^{(2)}\sim\hat{a}^\dag(\hat{a}^\dag\hat{a}+\hat{b}^\dag\hat{b}-N)$
&
\makecell[c]{$\sum_{n=0}^{N-m} f_{n}|n\rangle\otimes|m\rangle$}
&
$\makecell[c]{|N-m\rangle\otimes |m\rangle}$ \\
\hline

(3)
&
$\hat{L}_{1,-}^{(2)}\sim\hat{a}(\hat{a}^\dag\hat{a}+\hat{b}^\dag\hat{b}-N)$
&
\makecell[c]{$\sum_{n\geq N}^{\infty} f_{n}|n\rangle\otimes|m\rangle$}
&
$\makecell[c]{|N-m\rangle\otimes |m\rangle}$ \\
\hline

(4)
&
\makecell[c]{$\hat{L}_{1}^{(2)}\sim(\hat{a}^\dag+\hat{a})(\hat{a}^\dag\hat{a}+\hat{b}^\dag\hat{b}-N)$}
&
$\sum_{n=0}^\infty f_{n}|n\rangle\otimes|m\rangle$
&
$|N-m\rangle\otimes|m\rangle$\\
\hline

(5)
&
\makecell[c]{$\hat{L}_{1,+}^{(2)}\sim\hat{a}^\dag(\hat{a}^\dag\hat{a}+\hat{b}^\dag\hat{b}-N)$, \\ and \ $
\hat{L}_{1,-}^{(2)}\sim\hat{a}(\hat{a}^\dag\hat{a}+\hat{b}^\dag\hat{b}-N)$}
&
$\sum_{n=0}^\infty f_{n}|n\rangle\otimes|m\rangle$
&
$|N-m\rangle\otimes|m\rangle$\\
\hline

(6)
&
\makecell[c]{$ \hat{L}_{1,+}^{(2)}\sim\hat{a}^\dag(\hat{a}^\dag\hat{a}+\hat{b}^\dag\hat{b}-N)$, \\ and \  $\hat{L}_{2,\pm\xi}^{(2)}\sim \hat{a}\pm\hat{b}$}
&
\makecell[c]{$\sum_{n,m}f_{n,m}|n\rangle\otimes|m\rangle$ \\ for $n+m\leq N$}
&
$|N,\pm\xi\rangle$ \\
\hline

(7)
&
\makecell[c]{$\hat{L}_{1,+}^{(2)}\sim\hat{a}^\dag(\hat{a}^\dag\hat{a}+\hat{b}^\dag\hat{b}-N)$, \\ and \ $\hat{L}_2^{(2)}\sim \hat{a}^2\pm\hat{b}^2$}
&
\makecell[c]{$\sum_{n,m}f_{n,m}|n\rangle\otimes|m\rangle$ \\ for $n+m\leq N$ \\ $m \in\{\text{even}, \text{odd}\}$}
&
$\sim |N,\xi\rangle+(-1)^{N-m}|N,-\xi\rangle$ \\
\hline

(8)
&
\makecell[c]{$\hat{L}_{1,+}^{(d)}\sim\hat{a}^\dag(\hat{J}_c^{(d)}-N)$,\\ and \ $\hat{L}_{2\leq j \leq d,\xi_\bullet}^{(d)}=c_{j-1}\hat{a}_{j}-c_{j}\hat{a}_{j-1}$}
&
\makecell[c]{$\sum_{\{n_j\}}f_{\{n_j\}}|n_1\rangle\otimes\dots\otimes|n_d\rangle $ \\ for $n_1+\dots+n_d\leq N$}
&
\makecell[c]{$\sim |N,\xi_1,\dots,\xi_{d-1}\rangle$} \\
\hline

(9)
&
\makecell[c]{$\hat{L}_{1,+}^{(d)}\sim\hat{a}^\dag(\hat{J}_c^{(d)}-N)$,\\ and \ $\hat{L}_{2,\dots,d}^{(d)}=\begin{cases}
      c_{j-1}\hat{a}_{j}-c_{j}\hat{a}_{j-1} \quad \text{for }2\leq j\leq d-1,\\
      c_{j-1}^2\hat{a}_{j}^2-c_{j}^2\hat{a}_{j-1}^2 \quad \text{for } j=d,
    \end{cases}$}
&
\makecell[c]{$\sum_{\{n_j\}}f_{\{n_j\}}|n_1\rangle\otimes\dots\otimes|n_d\rangle $ \\ for $n_1+\dots+n_d\leq N$ \\ $n_d \in\{\text{even}, \text{odd}\}$}
&
\makecell[c]{$\sim |N,\xi_1,\dots,\xi_{d-1}\rangle$ \\ $+(-1)^{N-n_d}|N,-\xi_1,\dots,-\xi_{d-1}\rangle$} \\
\hline
\end{tabular}
\caption{Summary of the jump operators and their corresponding dissipatively prepared states discussed in section~\ref{sec_preparation}. The general initial states listed in the second column are intended to characterize the broad class of states that can be stabilized by the proposed dissipative protocols. In all cases except case (3), initialization in the vacuum state is sufficient, i.e., $n=m=n_j=0$. The single-mode counterparts of cases (2) and (3), i.e., $\hat{L}_{1,+}^{(1)}\sim \hat{a}^\dag(\hat{a}^\dag\hat{a}-N)$, , $\hat{L}_{1,-}^{(1)}\sim \hat{a}(\hat{a}^\dag\hat{a}-N)$, can similarly steer the system towards the target Fock state $|N\rangle$. The single dissipative channel generated by case (4) is decomposed into the two independent channels shown in (5). This decomposition is motivated by practical implementation constraints: since the simultaneous realization of the terms $\hat{a}^\dagger$ and $\hat{a}$ in (4) may require stringent frequency-matching conditions, whereas their separate implementation as distinct dissipative channels can be more feasible experimentally. Although the dynamics generated by (4) and (5) are not strictly equal, both schemes steer the same class of initial states toward an identical steady-state manifold. The constraints $m, n_d \in \{\mathrm{even}, \mathrm{odd}\}$ in (7) and (9) indicate that the corresponding summation indices are restricted to either even or odd integers. The parity factor $(-1)^{N-n_d}$ of the resulting cat states additionally depends on the parity of the initial excitation number $n_d$ in the last mode, since the engineered dissipation preserves the parity of that mode, as discussed below Eq.~\eqref{eq_su2_final_parity}. From (9), it follows that the maximum number of nonlinear wave-mixing processes required for the preparation and stabilization of an arbitrary $d$-mode multinomial cat state is five (comprising a four-wave-mixing jump operator and one auxiliary lossy mode), as determined by $\hat{L}_1^{(d)}$. Increasing the Lie algebra dimension $d$ only increases the number of jump operators, i.e., another linear combination $\hat{L}_d^{(d+1)}=c_{d-1}\hat{a}_d-c_d\hat{a}_{d-1}$, without increasing the maximum nonlinear order of wave-mixing required.} \label{table_summary_jump_operators}
\end{table*}

\section{Application}
While engineered dissipative dynamics provide a powerful route for state preparation, their role extends far beyond this. Due to their inherently stabilizing nature, engineered dissipation may also be viewed as a form of autonomous error correction mechanism, which continuously steers the system back to the target state subspace in the presence of unwanted noise~\cite{AlbertQST19b}. This property can be exploited to suppress loss and decoherence, thereby extending the effective lifetime of the target states. In the following, we show that this enhanced robustness manifests differently depending on the application: as increased coherence and measurement times in quantum metrology, and as prolonged bit-flip and phase-flip lifetimes in quantum computing.

Before discussing these applications, it is convenient to introduce a logical basis for the subspace spanned by two-mode binomial cat states \begin{gather}
    \nonumber |q_0\rangle=|N,-1\rangle, \qquad |q_1\rangle=|N,1\rangle,\\
    |\psi_\pm\rangle=\frac{1}{\sqrt{2}}(|q_0\rangle\pm|q_1\rangle),\qquad|\psi_{\pm,i}\rangle
    =\frac{1}{\sqrt{2}}(|q_0\rangle\pm i|q_1\rangle). \label{eq_identification}
\end{gather}
These states naturally form the vertices of an effective Bloch sphere associated with the two-dimensional manifold spanned by $|q_0\rangle$ and $|q_1\rangle$. Within this subspace, we can define the corresponding Pauli operators as \begin{align}
    \nonumber \hat{\sigma}_x&=|q_0\rangle\langle q_1|+|q_1\rangle\langle q_0|,\\
    \nonumber \hat{\sigma}_y&=-i|q_0\rangle\langle q_1|+i|q_1\rangle\langle q_0|,\\
    \hat{\sigma}_z&=|q_1\rangle\langle q_1|-|q_0\rangle\langle q_0|. \label{eq_pauli}
\end{align}
\begin{figure*}
    \centering
    \includegraphics[width=2\columnwidth]{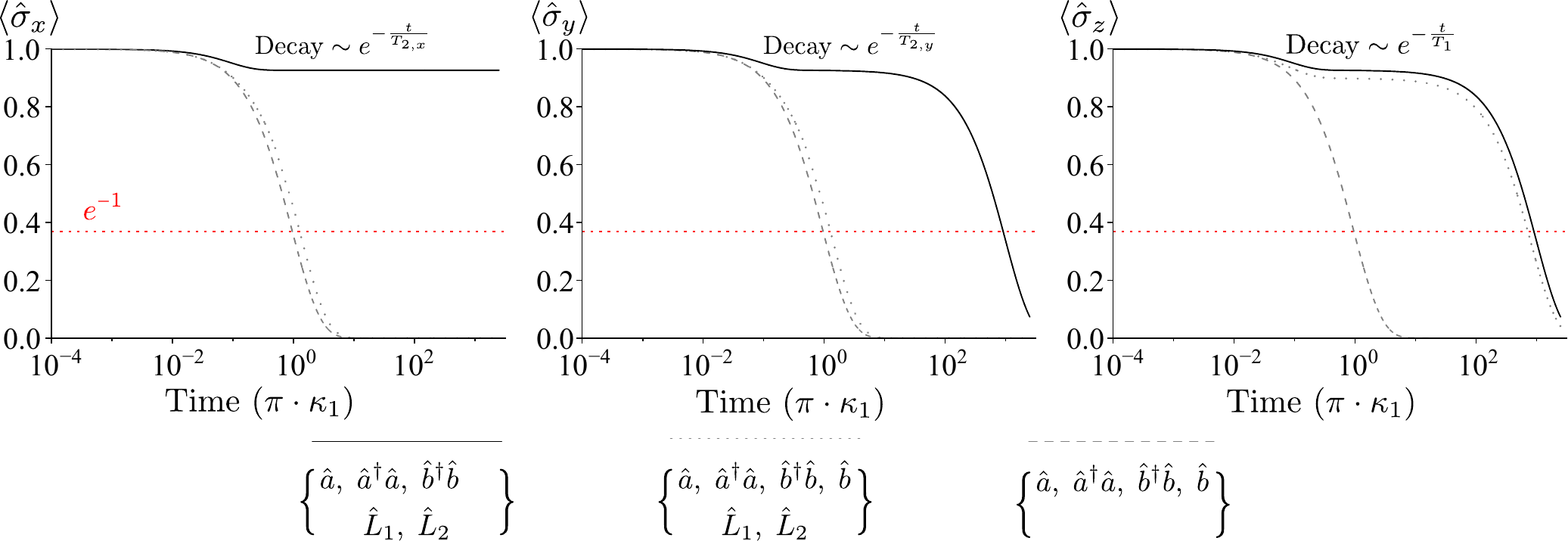}
    \caption{Decay of $\langle\hat{\sigma}_x\rangle\sim e^{-\frac{t}{T_{2,x}}}$, $\langle\hat{\sigma}_y\rangle\sim e^{-\frac{t}{T_{2,y}}}$, and $\langle\hat{\sigma}_z\rangle\sim e^{-\frac{t}{T_1}}$ with respect to the initial states $|\psi_+\rangle$, $|\psi_{+,i}\rangle$, and $|q_1\rangle$. The parameters used are $N=8$, $\hat{H}=0$, $\kappa_1=\kappa_2=\kappa=1$ for the engineered dissipation channels $\hat{L}_{1,+}^{(2)}$ and $\hat{L}_2^{(2)}$, and $\gamma_a=\gamma_b=\gamma_\phi=0.03\kappa$ for the error sources $\hat{a}$, $\hat{b}$, $\hat{a}^\dag\hat{a}$ and $\hat{b}^\dag\hat{b}$. The red dotted line marks the value of $e^{-1}$ to indicate the $T_1$ and $T_2$ times. The dashed line shows the exponential decays of $\langle\sigma_x\rangle$, $\langle\sigma_y\rangle$, and $\langle\sigma_z\rangle$ in the absence of $\hat{L}_{1,+}^{(2)}$ and $\hat{L}_2^{(2)}$. The solid line shows how the $x$-component of coherence time $T_{2,x}$ can be extended to infinity, as well as how $T_{2,y}$ and $T_1$ can be increased by two to three orders of magnitude. The dash-dotted line shows that $\hat{L}_{1,+}^{(2)}$ and $\hat{L}_2^{(2)}$ cannot fight against the additional error channel $\hat{b}$ to prolong the $T_2$ time, but $T_1$.}
    \label{fig_decay_xyz}
\end{figure*}
The expectation values of the effective Pauli operators $\hat{\sigma}_{x,y,z}$ evaluated on the logical initial states $|\psi_+\rangle$, $|\psi_{+,i}\rangle$, and $|q_1\rangle$, provide a convenient way to track decoherence within the target state manifold. In the presence of noise, these observables decay as $\langle\hat{\sigma}_x\rangle \sim e^{-\frac{t}{T_{2,x}}}$, $\langle\hat{\sigma}_y\rangle \sim e^{-\frac{t}{T_{2,y}}}$, and $\langle\hat{\sigma}_z\rangle \sim e^{-\frac{t}{T_1}}$, which in turn define the relaxation time $T_1$ and direction-dependent dephasing times $T_{2,x}$ and $T_{2,y}$ of the system.

From Fig.~\ref{fig_decay_xyz}, we observe that in the absence of engineered dissipation ($\hat{L}_{1,+}^{(2)}$, $\hat{L}_2^{(2)}$), all decay channels are governed by the intrinsic noise rates $\sim 1/\gamma_\bullet$, and the Bloch vector in the $xy$-plane contracts isotropically. In this regime, it is sufficient to monitor only one of the $x$- or $y$-component of the Bloch vector, since a uniform decay in both components implies \begin{align}
    \nonumber \norm{v_{xy}(t)}&=\sqrt{\norm{v_x(0) e^{-\frac{t}{T_2}}}^2+\norm{v_y(0)e^{-\frac{t}{T_2}}}^2},\\
    &=\norm{v_{xy}(0)}e^{-\frac{t}{T_2}}.
\end{align}
However, this symmetry breaks once engineered dissipation channels are included (excluding the single-photon loss $\sqrt{\gamma_b}\hat{b}$). As illustrated in Fig.~\ref{fig_decay_xyz}, the $x$-component becomes strongly protected by the dissipative stabilization and remains constant close to unity $\langle\hat{\sigma}_x\rangle\sim 0.9$ over time. In contrast, the $y$-component keeps decaying, but on significantly extended timescales compared to the bare system without engineered dissipation. This symmetry breaking can be interpreted as a form of anisotropic transverse decoherence, which may itself be useful as a diagnostic tool for identifying dominant noise mechanisms in the system~\cite{RowerPRA25, ChoinQI22}. From a more structural viewpoint, this behavior originates from the fact that symmetry operations on the logical Bloch sphere generally do not commute with the combined action of unwanted noise and engineered dissipation \begin{align}
    \nonumber \langle\hat{\sigma}_x\rangle=\mathrm{Tr}(\hat{\sigma}_y\hat{\mathcal{R}}\hat{\mathcal{S}}\hat{\rho}_{+}),\qquad &\langle\hat{\sigma}_y\rangle=\mathrm{Tr}(\hat{\sigma}_y\hat{\mathcal{S}}\hat{\mathcal{R}}\hat{\rho}_{+}),\\
    \text{hence}\quad \langle\hat{\sigma}_x\rangle\neq\langle\hat{\sigma}_y\rangle \iff& [\hat{\mathcal{R}},\hat{\mathcal{S}}]\neq 0,
\end{align}
where $\hat{\rho}_+=|\psi_+\rangle\langle\psi_+|$, $\hat{\mathcal{S}}$ denotes the open-system evolution, and $\hat{\mathcal{R}}$ is a unitary rotation by $\frac{\pi}{2}$ around the $z$ axis of the logical Bloch sphere.

Nevertheless, the protection offered by engineered dissipation is not universal with respect to all noise channels. In particular, the inclusion of single-photon loss $\sqrt{\gamma_b}\hat{b}$ introduces a phase-flip process within the target state (or logical) manifold, which cannot be corrected by the engineered dissipators. As a result, both $\langle\hat{\sigma}_x\rangle$ and $\langle\hat{\sigma}_y\rangle$ become sensitive to this error channel, implying that the two-mode binomial cat states remain susceptible to phase-flip errors. More generally, this behavior is characteristic of dissipatively stabilized cat states, where the bit-flip error rate (associated with the $z$ component of the Bloch vector) is exponentially suppressed at the expense of a linearly enhanced phase-flip error rate (associated with the $x$ and $y$ components) \cite{LescanneNP20, MirrahimiNJP14a, A&B}. Remarkably, our engineered dissipation circumvents this trade-off. As shown in Fig.~\ref{fig_decay_xyz}, the exponential enhancement of the bit-flip lifetime is achieved without any observable reduction of the phase-flip lifetime compared to the bare noisy case, demonstrating that the improved protection against bit-flip errors does not come at the cost of an amplified phase-flip susceptibility.

\subsection{Quantum metrology}\label{sec_metro}
Although the notion of the cat state as a superposition of macroscopically distinct quantum states is physically vague and subjectively defined, a useful quantitative characterization can be given in terms of the so-called q-index ($1\le q\le 2$)~\cite{ShimizuPRL05}. Within this framework, cat states with $q=2$ are particularly valuable as probing states for quantum metrology, as they enable Heisenberg-limited sensitivity in closed system Ramsey protocols~\cite{TatsutaPRA19a}. In what follows, we show that the two-mode binomial cat states $|\psi_\pm\rangle$ introduced in Eq.~\eqref{eq_def_cat_state} belong to this class by demonstrating that they satisfy $q=2$. We then discuss how the engineered dissipative dynamics in Eq.~\eqref{Eq02} can preserve this scaling by effectively prolonging the relevant coherence time $T_2$, thereby enabling near-Heisenberg-limited sensitivity even in the presence of environmental noise.

First, we note that the q-index is defined through the scaling behavior of the second order function, \begin{align}
    \mathrm{max}\left\{N,\operatorname*{\mathrm{max}}_{\hat{A},\hat{\eta}}\left\{\mathrm{Tr}\left(\hat{\rho}[\hat{A},[\hat{A},\hat{\eta}]]\right)\right\}\right\}&\sim\mathcal{O}(N^q), \label{Eq07}
\end{align}
where $\hat{\rho}$ is the probe state, $\hat{A}$ is the measured observable whose operator norm is assumed to be $\|\hat{A}\|\sim N$ (or equivalently, a two-wave mixing operator), and $\hat{\eta}$ denotes the corresponding projective measurement operator such that $\hat{\eta}^2=\hat{\eta}$ \cite{ShimizuPRL05}. While mathematically, Eq.~\eqref{Eq07} does not restrict the choice of $\hat{A}$ and $\hat{\eta}$, in practice these operators are preferably selected according to the available experimental resources. Specifically, $\hat{A}$ is typically determined by the accessible interaction Hamiltonians in the platform, while $\hat{\eta}$ is chosen according to feasible readout schemes. In addition, to cope with the engineered dissipation to stabilize the probing state manifold, $\hat{A}$ may also be chosen such that it does not push the population out of this manifold.

To give a more concrete example, consider the task of detecting a weak magnetic flux $\Phi_\mathrm{ext}$ in a superconducting set-up consisting of two resonators coupled via a beam-splitter interaction
\begin{align}
    \hat{H}_\mathrm{int}=\frac{\theta(\Phi_\mathrm{ext})}{2}(\hat{a}^\dag\hat{b}+\hat{a}\hat{b}^\dag), \label{eq_bs_int}
\end{align}
which can be as well implemented using the ATS nonlinear element. In this setting, the coupling strength $\theta(\Phi_\mathrm{ext})$ depends on the external flux across the nonlinear element, and therefore measuring the beam-splitter interaction strength provides us a direct estimate of the weak magnetic flux. In the $q$-index formalism depicted by Eq.~\eqref{Eq07}, such measurement set-up corresponds to identifying the following ingredients: $\hat{\rho}=\hat{\eta}=|\psi_\pm\rangle\langle\psi_\pm|$, and $\hat{A}=\frac{1}{2}(\hat{a}^\dag\hat{b}+\hat{a}\hat{b}^\dag)$. Here, we stress that the choice of the read-out projector $\hat{\eta}$ to measure the beam-splitter strength is not unique, and the specific choice of measurement is primarily motivated by experimental accessibility. In practice, if necessary, this read-out scheme may be coupled to the $a$-mode instead of the $b$-mode. Although this introduces additional loss and decoherence to the $a$-mode, but as discussed previously, the engineered dissipation can stabilize the probing states within the steady state manifold against noise channels acting on the $a$-mode. In this work, we adopt $\hat{\eta}=|\psi_\pm\rangle\langle\psi_\pm|$ because it provides both a transparent way to verify the $q=2$ scaling in Eq.~\eqref{Eq07}, and is also closely related to parity measurements over the even and odd excitation sectors $\{|2n, N-2n\rangle, |2n+1, N-2n-1\rangle\}$ \cite{TeohPNAS23, GoviaPRA15, SunN14}. With this choice, Eq.~\eqref{Eq07} exhibits a quadratic scaling $\mathcal{O}(N^2)$, confirming that the $\mathfrak{su}(2)$ cat states in Eq.~\eqref{eq_def_cat_state} correspond to $q=2$. The former follows from the fact that the basis states $|q_{1,0}\rangle$ are eigenstates of $\hat{A}$ with eigenvalues $\pm\frac{N}{2}$ respectively.

Having identified the $q=2$ scaling, we have shown that two-mode binomial cat states naturally support Ramsey-type metrology in Heisenberg's limit. The achievable precision can be quantified using the normalized sensitivity~\cite{TatsutaPRA19a}
\begin{align}
    \Delta\theta\sqrt{T}=\frac{\sqrt{P(1-P)}}{\left|\frac{dP}{d\theta}\right|}\sqrt{t_\mathrm{int}}, \label{def_sensitivity}
\end{align}
where $P=\mathrm{Tr}\{\hat{\rho}(t_\mathrm{int})\hat{\eta}\}$, $t_\mathrm{int}$ is the interrogation time for each run, and $T$ is the total measurement time. Here, we note that the smaller the value of $\Delta\theta\sqrt{T}$, the more sensitive the measurement is.

\begin{figure}
    \centering
    \includegraphics[width=\linewidth]{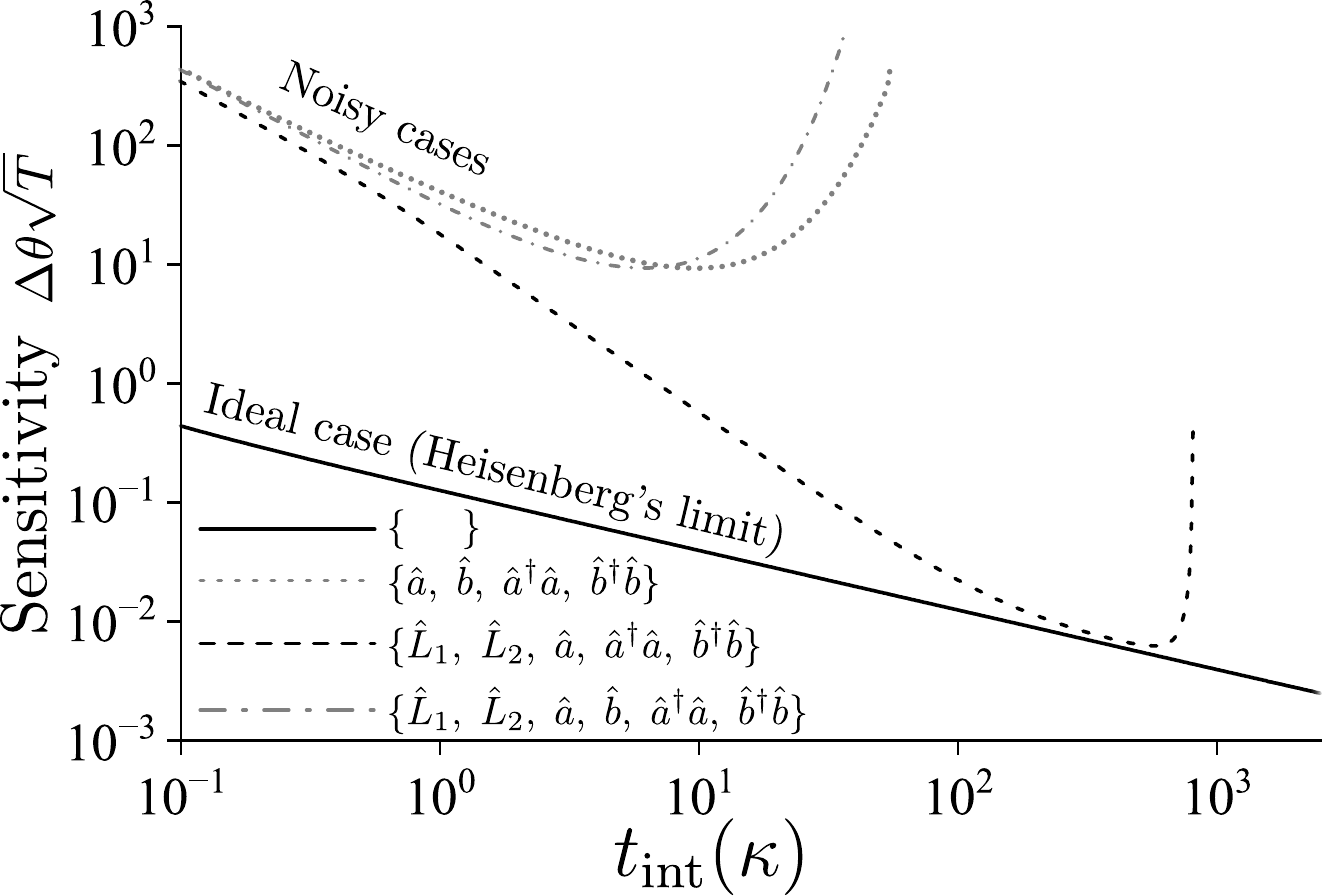}
    \caption{Open-system simulation of the normalized sensitivity defined in Eq.~\eqref{def_sensitivity} using the even two-binomial cat state $|\psi_+\rangle$. The parameters are chosen as $N=8$ and $\kappa_1=\kappa_2=\kappa=1$ for the engineered dissipation channels $\hat{L}_{1,+}^{(2)}$ and $\hat{L}_2^{(2)}$. The noise processes are characterized by $\gamma_a=\gamma_b=\gamma_\phi=0.03\kappa$, corresponding to single-photon loss in modes $a$ and $b$, as well as dephasing in both modes. The parameter (very weak magnetic flux) to be estimated is set to $\theta=0.0005\kappa$.}
    \label{Fig_05}
\end{figure}
However, in realistic scenarios, decoherence fundamentally limits the performance of such protocols. That is, unwanted noise channels will reduce the coherence time $T_2$ of the probing state, thereby restricting the available interrogation time $t_\mathrm{int}$ over which the system can accumulate information about the parameter $\theta$. This limitation is directly reflected in the sensitivity defined in Eq.~\eqref{def_sensitivity}. As the system evolves under noise, it rapidly approaches to a steady state, causing $P$ to become constant, and thus effectively independent of $\theta$. Consequently, the derivative in the denominator vanishes $|\frac{dP}{d\theta}|\to 0$, which in turn leads to a divergence of the sensitivity $\Delta\theta\sqrt{T}$. In other words, beyond a certain interrogation time, additional evolution no longer improves the measurement precision but instead degrades it. This sets a fundamental bound on the achievable sensitivity, preventing the system from reaching the Heisenberg limit in the presence of noise~\cite{ReiterNC17}. This U-turn behavior is illustrated in Fig.~\ref{Fig_05}, where in the presence of noise, the sensitivity initially decreases towards the Heisenberg limit but eventually reaches a minimum, after which it reverses the trend and increases with time. In Fig.~\ref{Fig_05}, we observe that in the absence of the engineered dissipation (dotted line) or in the presence of single-photon loss in the $b$-mode (dash-dotted line), the probe state (two-mode binomial cat state) rapidly loses coherence. As a result, the optimal interrogation time is severely limited, causing the sensitivity curve to exhibit a U-turn long before approaching Heisenberg's limit. In contrast, when the single-photon loss channel $\sqrt{\gamma_b}\hat{b}$ is absent (dashed line), the engineered dissipation effectively confines the dynamics to the probe state manifold, and suppresses leakage induced by other noise channels. This effectively prolongs the coherence time of the probe state, allowing the interrogation time to increase until the sensitivity approaches to Heisenberg's limit.

The origin of this behavior can be understood back in the Bloch-sphere language presented in Fig.~\ref{fig_decay_xyz}. The maximum allowed interrogation time is ultimately determined by the coherence time $T_2$, which characterizes the decay of the transverse Bloch-vector components ($x$- and $y$-components) according to \begin{align}
    \langle\psi_+|\hat{\sigma}_x|\psi_+\rangle &\sim e^{-t/T_{2,x}},\qquad \langle\psi_{+,i}|\hat{\sigma}_y|\psi_{+,i}\rangle \sim e^{-t/T_{2,y}}.
\end{align}
As discussed previously, the engineered dissipation produces a pronounced asymmetry between the decoherence channels. In the absence of single-photon loss, $\hat{L}_{1,+}^{(2)}$ and $\hat{L}_2^{(2)}$ stabilize the $x$-component to a steady state very close to $|\psi_+\rangle$, which effectively extends $T_{2,x}$ to infinity. Although the $y$-component continues to decay, its coherence time is extended by several orders of magnitude, i.e., for the parameters considered in Fig.~\ref{fig_decay_xyz}, $T_{2,y}$ exceeds the bare coherence time by roughly two to three orders of magnitude. Consequently, the available interrogation time is significantly increased, leading directly to an enhanced metrological sensitivity. This stabilization mechanism enables sensitivities that remain close to Heisenberg's limit even in the presence of unwanted noise channels.

\subsection{Quantum computing}\label{sec_quan_computing}
Beyond their role in quantum metrology, two-mode binomial cat states can also be naturally employed for quantum information processing. In this context, the two components $|N,\pm1\rangle$ define a logical qubit basis $|q_{0,1}\rangle$ introduced in Eq.~\eqref{eq_identification}. A general qubit state can thus be encoded as \begin{align}
    |\varphi\rangle \sim |q_0\rangle + e^{i\phi}|q_1\rangle,
\end{align}
with the symmetric and antisymmetric combinations corresponding to the even and odd two-mode binomial cat states $|\psi_\pm\rangle$. This encoding realizes a two-mode bosonic qubit, which has the advantage of being robust against noise. In particular, it has been shown that multi-mode cat qubits can exhibit enhanced error-correction capabilities as the number of modes increases~\cite{AlbertQST19b}, which naturally motivates the use of two-mode binomial cat states, as well as its extensions of $d$-mode multinomial cat states. Moreover, since these states involve finite superpositions of Fock states, they can exactly correct higher-order dephasing errors, which is in contrast to that of infinite superpositions where only approximate correction is possible \cite{AlbertQST19b, AlbertPRA18}. Alternatively, this may also be seen as a consequence of the overlap between non-compact coherent states, such as $\langle\alpha|-\alpha\rangle=e^{-2|\alpha|^2}$, being essentially an infinite series that can only asymptotically converge to zero, but never reach zero. In this section, we show that the engineered dissipation introduced in Eq.~\eqref{Eq02} can autonomously stabilize the cat qubits encoded using the two-mode binomial cat states. In close analogy to previously proposed schemes~\cite{AlbertQST19b, GertlerPQ23a}, the dissipative dynamics suppress bit-flip errors by confining the system within the logical subspace, while also partially mitigating phase-flip errors. We find, that although the encoding involves two modes, which is, in principle, exposed to a larger set of noise channels, the presence of engineered dissipation effectively suppresses all of them except of one. What remains is only the single-photon loss in one mode, akin to the case of a single-mode cat state. This leaves the two-mode binomial code sensitive to only a single error channel, while the remaining channels being strongly suppressed.

In the context of quantum computing we use bit- and phase-flip times to quantify the performance of the encoded qubit. Using the Pauli operators defined in Eq.~\eqref{eq_pauli}, these times can be extracted from the decay of expectation values as
\begin{align}
    \langle q_{0,1}|\hat{\sigma}_z|q_{0,1} \rangle \sim e^{-t/T_{\text{bit-flip}}}, \quad
    \langle \psi_\pm|\hat{\sigma}_x|\psi_\pm \rangle \sim e^{-t/T_{\text{phase-flip}}},
\end{align}
following~\cite{HillmannPRA23, LescanneNP20}. We note that the phase-flip time $T_{\text{phase-flip}}$ coincides with the previously discussed coherence time $T_{2,x}$ (as well as $T_{2,y}$), whose behavior has already been presented in Fig.~\ref{fig_decay_xyz}(a) and (b). It therefore remains to examine the bit-flip dynamics shown in Fig.~\ref{fig_decay_xyz}(c). We observe that the engineered dissipation $\hat{L}_{1,+}^{(2)}$ and $\hat{L}_2^{(2)}$ extend the bit-flip time by two to three orders of magnitude in the presence of all the dominant error channels $\{\hat{a},\, \hat{a}^\dagger \hat{a},\, \hat{b}^\dagger \hat{b},\, \hat{b}\}$. This robustness can be understood from the fact that bit-flip errors depend only on the populations of $|q_0\rangle$ and $|q_1\rangle$, rather than their relative phase. In particular, although single-photon loss in the $b$ mode induces a phase-flip between the cat components, it does not alter their populations and therefore does not directly contribute to bit-flip errors.

For completeness, we consider the corresponding dynamics for Schr\"odinger's cat states constructed from the single-mode coherent states $|\pm\alpha\rangle$. In this case, the dominant noise channels are $\{\hat{a},\, \hat{a}^\dagger \hat{a}\}$, and the engineered dissipation $\hat{L}=\hat{a}^2-\alpha^2$ primarily suppresses bit-flip errors, as reflected in the decay of $\langle \hat{\sigma}_z \rangle$. However, phase-flip errors captured by $\langle \hat{\sigma}_x \rangle$ and $\langle \hat{\sigma}_y \rangle$ remain essentially uncorrected, due to the same reasons as that of the two-mode binomial code. This situation mirrors the two-mode case in a complementary way: although the two-mode binomial encoding is exposed to a larger set of noise channels, the engineered dissipation effectively reduces the relevant error mechanism to a single dominant channel. In contrast, the single-mode encoding involves fewer physical noise channels, but the fragile noise channel is again governed by a single dominant phase-flip process. In both cases, the effectiveness of the dissipative protection is ultimately limited by one leading error channel, regardless of different physical origins.

\section{Conclusion}
In conclusion, we have developed a dissipative engineering framework for the autonomous preparation and stabilization of compact Lie-algebraic cat states in bosonic systems. As a concrete example, we focused on two-mode binomial cat states associated with the $\mathfrak{su}(2)$ algebra. These states are distinct from Schr\"odinger cat states and pair-cat states, which are associated with the non-compact $\mathfrak{h}(1)$ and $\mathfrak{su}(1,1)$ algebras respectively. Since compact cat states have finite support in the number basis, they cannot be stabilized by the usual ladder operator eigenvalue treatment. Their stabilization therefore requires a different dissipative design principle.

Our construction uses several engineered dissipative channels to enforce the defining properties of the target state. For the two-mode binomial cat state, one dissipator fixes the total excitation-number manifold, while a second fixes the relative structure between the two modes. Together, these processes steer the system from the vacuum towards the target cat state, and stabilize it against unwanted photon loss and dephasing channels. We further showed that the same stabilization principle can extend to $d$-mode multinomial cat states associated with higher-dimensional $\mathfrak{su}(d>2)$ algebras, providing a scalable route to compact multimode cat states.

We also proposed an experimentally accessible implementation in superconducting circuits for the designed dissipative processes. The proposed realization consists of coupled resonators, lossy auxiliary modes, and a nonlinear element such as ATS or SNAIL coupler. This places the preparation and stabilization scheme within current directions in superconducting reservoir engineering, making it relevant to state-of-the-art bosonic quantum platforms.

Finally, we showed that the proposed stabilization mechanism is useful not only for state preparation, but also for quantum technologies. For example, in quantum metrology, the engineered dissipation confines the two-mode binomial cat states to the probe state manifold, thereby prolonging the optimal interrogation time and preserving the Heisenberg limited scaling even in the presence of noise. In quantum information processing, the same cat state (or probe state) manifold defines a protected bosonic encoding with substantially extended coherence and logical lifetimes. Our results therefore establish compact $\mathfrak{su}(d\geq2)$ cat states as a distinct class of dissipatively stabilizable nonclassical resources, providing a route towards their realization and application in (multimode) quantum technologies.

\section{Acknowledgements}
We thank Yiwen Chu and Alexander Grimm for helpful discussions.
We acknowledge funding by the Deutsche Forschungsgemeinschaft through the Emmy Noether program (Grant No.~ME 4863/1-1). 

\balance
\clearpage
\appendix

\section{Derivation of the \texorpdfstring{$\mathfrak{su}(2)$}{} coherent states} \label{app_01}
Here, we  briefly revisit the explicit expression of a
$\mathfrak{su}(2)$ coherent state in the Fock basis.
Following \cite{Perelomov86}, a general $SU(2)$ group element in their operator representation is called the displacement operator $\hat{D}(\xi)$. This displacement operator can be normal ordered as 
\begin{align}
    \nonumber \hat{D}(\xi)&=e^{\tan^{-1}(\xi)(\hat{a}^\dag\hat{b}-\hat{a}\hat{b}^\dag)} \\
    &=e^{\xi \hat{a}^\dag\hat{b}}e^{\frac{1}{2}\ln(1+|\xi|^2)(\hat{a}^\dag\hat{a}-\hat{b}^\dag\hat{b})} e^{-\xi \hat{a}\hat{b}^\dag}. \label{eq_su2_disp_normal_order}
\end{align}
Using Eq.~\eqref{eq_su2_disp_normal_order}, we obtain the $\mathfrak{su}(2)$ coherent state as
\begin{align}
    \nonumber |N,\xi\rangle&=\hat{D}(\xi)|0,N\rangle
    \\
        \nonumber &
        =e^{\xi \hat{a}^\dag\hat{b}}e^{\frac{1}{2}\ln(1+|\xi|^2)(\hat{a}^\dag\hat{a}-\hat{b}^\dag\hat{b})} e^{-\xi \hat{a}\hat{b}^\dag}|0,N\rangle
    \\
    \nonumber &=
    \frac{1}{\sqrt{(1+|\xi|^2)^{N}}}\sum_{n=0}^\infty  \xi^n 
    \sqrt{\frac{N!}{n!(N-n)!}}|n,N-n\rangle\\\
    &=\frac{1}{\sqrt{(1+|\xi|^2)^{N}}} \sum_{n=0}^\infty \binom{N}{n}^{\frac{1}{2}} \xi^n |n,N-n\rangle.
    \label{eq_def_2m_binomial}
\end{align}
Since we work in the two-mode bosonic (or Schwinger) representation, in which the displacement operator Eq.~\eqref{eq_su2_disp_normal_order} is expressed in terms of bosonic creation/annihilation operators, and the system state is written as the tensor product between two bosonic resonator states, e.g., $|n,N-n\rangle\equiv |n\rangle\otimes|N-n\rangle$, such $\mathfrak{su}(2)$ coherent states are also known as the two-mode binomial coherent states. Here, the two-mode nature originates to the effective two-mode bosonic representation, and the term ``binomial" describes the underlying binomial distribution $\sim\binom{N}{n}\equiv\frac{N!}{n!(N-n)!}$ of the number states $|n, N-n\rangle$ in Eq.~\eqref{eq_def_2m_binomial}.

\section{q-index of two-mode binomial cat states}\label{app_05} 
Here we show how the  $q$-index defined as
\begin{align}
    \mathrm{max}\left\{N,\operatorname*{\mathrm{max}}_{\hat{A},\hat{\eta}}\left\{\mathrm{Tr}\left(\hat{\rho}[\hat{A},[\hat{A},\hat{\eta}]]\right)\right\}\right\}&\sim\mathcal{O}(N^q), 
    \label{Eq07App}
\end{align}
takes indeed the value of $q=2$ for two-mode binominal cat states. Just as in the main text  we choose  the observable $\hat{A}=\frac{1}{2}(\hat{a}^\dag\hat{b}+\hat{a}\hat{b}^\dag)$, the probing state $\hat{\rho}=|\psi_\pm\rangle\langle\psi_\pm|$, and the projective measurement $\hat{\eta}=\hat{\rho}$. 
The binominal states are eigenstates to the observable $\hat{A}$, i.e., $\hat{A}|N,\pm1\rangle= \pm \frac{N}{2}|N,\pm 1\rangle$, thus
 acting with the observable $\hat{A}$ onto
 the even/odd two-mode binomial cat states $|\psi_\pm\rangle$ we obtain
\begin{align}
\hat{A} |\psi_\pm\rangle   =  \frac{N}{2} 
|\psi_\mp\rangle, \hspace{0.2cm}  \text{with} \hspace{0.2cm}  
|\psi_\pm\rangle\sim|N,-1\rangle\pm |N,1\rangle,
\end{align}
Hence, the double commutator in Eq.~\eqref{Eq07App} reduces to 
\begin{align}
    \nonumber [\hat{A},[\hat{A},\hat{\eta}]]&=\hat{A}^2\hat{\eta}-2\hat{A}\hat{\eta}\hat{A}+\hat{\eta}\hat{A}^2,\\
    &=\frac{N^2}{2}(|\psi_\pm\rangle\langle\psi_\pm|-|\psi_\mp\rangle\langle\psi_\mp|).
\end{align}
With this, we can evaluate the trace in Eq.~\eqref{Eq07App} as
\begin{align}
    \nonumber \mathrm{Tr}(\hat{\rho}[\hat{A},[\hat{A}\hat{\eta}]])&=\frac{N^2}{2}\langle\psi_\pm|(|\psi_\pm\rangle\langle\psi_\pm|-|\psi_\mp\rangle\langle\psi_\mp|)|\psi_\pm\rangle 
    =\frac{N^2}{2}. 
\end{align}
Eventually, the $q$-value for two-mode binomial cat states is encoded in \begin{align}
    \mathrm{max}\left\{N,\frac{N^2}{2}\right\}&=\mathcal{O}(N^2),
\end{align}
which indicates that $q=2$.

\section{Adiabatic elimination of the lossy mode}\label{app_02}
To engineer a non-trivial dissipative process $\hat{L}$ we utilize a lossy auxiliary mode. The Hilbert space of the combined system thus becomes $\mathcal{H} = \mathcal{H}_{C} \otimes \mathcal{H}_{L}$. We denote the auxiliary mode with annihilation operator $\hat{c}$, and assume a coupling Hamiltonian of the form $\hat{H}=g_c\hat{L}\hat{c}^\dag+h.c.$. In what follows, we illustrate how such a scenario results in a purely dissipative process with the jump operator $\sqrt{\kappa_a}\hat{L}$ acting on the reduced Hilbert space $\mathcal{H}_{L}$. The rate $\kappa_{a}$ is then determined via $\kappa_a= 4g_c^2/\kappa_c$, where $\kappa_{c}$ denotes the decay rate of the auxiliary mode. We follow here the derivation in Ref. \cite{Hauer2018}.

We start from the dynamics of the total systems density matrix $\hat{\sigma}$, which can be captured in the master equation
\begin{align}\label{Eq.masterequation}
 \frac{\partial}{\partial t} \hat{\sigma} =  \left\{    \hat{\mathcal{D}}_{C} +  \hat{\mathcal{D}}_{CL} \right\}\hat{\sigma},
\end{align}
with the superoperators
\begin{align}
 \hat{\mathcal{D}}_{C}   =&  \kappa_c
 \left[ \hat{c} \bullet \hat{c}^{\dag} - \frac{1}{2}
 \left\{ \hat{c}^{\dag} \hat{c} ,\bullet \right\}
 \right],
 \nonumber \\
 \hat{\mathcal{D}}_{CL}   =&- i \left[g_{c} \left[ \hat{c}^{\dag} \hat{L}   +   \hat{c}  \hat{L}^{\dag}  \right], \bullet \right] ,
\end{align}
where $\bullet$ denotes the operator the superoperator is acting upon. The superoperator $\hat{\mathcal{D}}_{C}$ acts only on the auxiliary-mode  space and $\hat{\mathcal{D}}_{CL}$ acts on the full Hilbert space. Note, we could as well have included further dissipative and coherent processes acting solely on $\mathcal{H}_{L}$, however, to simply our analysis we neglect those processes here.
We now move  into a new interaction picture
\begin{align}
 \hat{\sigma}^{\prime} = e^{- \hat{\mathcal{D}}_{C}   t } \hat{\sigma}
 \hspace{0.1cm} \Rightarrow \hspace{0.1cm}
 \frac{\partial}{\partial t}  \hat{\sigma}^{\prime} = e^{-  \hat{\mathcal{D}}_{C}   t } \hat{\mathcal{D}}_{CL} e^{  \hat{\mathcal{D}}_{C}   t } \hat{\sigma}^{\prime}
                                  \equiv 
\hat{\mathcal{D}}_{CL}^{\prime}  \hat{\sigma}^{\prime},
\end{align}
here we simply used the product rule and Eq.~(\ref{Eq.masterequation}) to derive the master equation for $\hat{\sigma}^{\prime}$.
This master equation can  formally be integrated, we obtain the solution
\begin{align}
 \hat{\sigma}^{\prime}(t) =& \hat{\sigma}^{\prime}(0) + \int\limits_{0}^{t} d\tau \; \hat{\mathcal{D}}_{CL}^{\prime}(\tau)  \hat{\sigma}^{\prime}(\tau),
\end{align}
which can be substituted back into the master equation and by additionally performing the trace over system $C$ we obtain the reduced master equation for the system $L$:
\begin{align}\label{Eq.masterequationIP}
  \frac{\partial}{\partial t}  \trA \left\{  \hat{\sigma}^{\prime}(t) \right\} 
  \equiv & \frac{\partial}{\partial t} \hat{\rho}  (t) = \trA \left\{ \hat{\mathcal{D}}_{CL}^{\prime}(t)  \hat{\sigma}^{\prime}(0) \right\}
       \nonumber \\  &
 + \int\limits_{0}^{t} d\tau \;
    \trA \left\{  \hat{\mathcal{D}}_{CL}^{\prime}(t)  \hat{\mathcal{D}}_{CL}^{\prime}(\tau)  \hat{\sigma}^{\prime}(\tau)\right\} .
\end{align}
Thus, we require dynamics of the superoperator
\begin{align}
\hat{\mathcal{D}}_{CL}'(t)   =&
 -i g_{c}
 \bigg\{ \hat{\mathcal{C}}_2^{\prime}(t)
     \hat{L} \bullet
            +
            \hat{\mathcal{C}}_1^{\prime}(t)
            \hat{L}^{\dag} \bullet
      -  \hat{\mathcal{C}}_1^{\prime\dag}(t)  \hat{L}
            -  \hat{\mathcal{C}}_2^{\prime \dag}(t)  \hat{L}^{\dag}
    \bigg\},
\end{align}
with the  definitions  
\begin{align}
\hspace{0.5cm} 
  \hat{\mathcal{C}}_1^{\prime}(t) =   e^{-  \hat{\mathcal{D}}_{C}  t } \hat{c}        \bullet e^{ \hat{\mathcal{D}}_{C}  t },
 \hspace{0.3cm} 
  \hat{\mathcal{C}}_2^{\prime}(t) =   e^{-  \hat{\mathcal{D}}_{C}  t } \hat{c}^{\dag}        \bullet e^{ \hat{\mathcal{D}}_{C}  t }.
\end{align}
To further evaluate the superoperator $\hat{\mathcal{D}}_{CL}^{\prime}  $ we need the
dynamics of the auxiliary mode superoperators in this interaction picture:
\begin{align}
   \hat{\mathcal{C}}_1^{\prime}(t) =&   \hat{c}        \bullet   e^{-   \frac{ \kappa_c}{2}  t},
   \hspace{0.5cm}
   \hat{\mathcal{C}}_2^{\prime}(t) =
   \hat{c}^{\dag}    \bullet
   e^{ \frac{\kappa_c }{2}   t}
-   \bullet \hat{c}^{\dag} \left[ e^{+  \frac{\kappa_c }{2}t} -   e^{-  \frac{\kappa_c }{2}t}\right].
\end{align} 
Similar to the derivation of a standard master equation, we assume from now on that the system $L$ and the auxiliary mode $C$ remain uncorrelated throughout the evolution. The total density matrix therefore factorizes as $\hat{\sigma}(t) = \hat{\rho}(t) \otimes \hat{\rho}_C(t)$, where $\hat{\rho}(t)$ denotes the reduced density matrix of the system and $\hat{\rho}_C(t)$ the density matrix of the auxiliary mode. We further make the Born approximation and assume that the state of the auxiliary mode is not affected by its coupling to the system, i.e., we set
$\hat{\rho}_C(t) \approx \hat{\rho}_C(0)$. Consequently, the total density matrix is approximated by $\hat{\sigma}(t) \approx \hat{\rho}(t) \otimes \hat{\rho}_C(0)$. For an auxiliary mode initially prepared in the vacuum state, this reduces to $\hat{\sigma}(t) \approx \hat{\rho}(t) \otimes |0\rangle\langle0|$. Under this assumption the first term in Eq.~(\ref{Eq.masterequationIP}) vanishes, and  the second term yields \begin{align}
 & \trA \left\{  \hat{\mathcal{D}}_{CL}^{\prime}(t) \hat{\mathcal{D}}_{CL}^{\prime}(\tau) \;  \hat{\rho}(\tau)   \otimes   |0 \rangle\langle 0|  \right\}
 \nonumber \\ &
 =
    g_{c}^2
          \left\{
                  \hat{L}  \bullet
                  \hat{L}^{\dag}
                     -
                  \hat{L}^{\dag}
                  \hat{L}  \bullet
          \right\}
          e^{ -   \frac{\kappa_c }{2}  (t - \tau) } \hat{\rho}(\tau) + h.c. \; .
 \end{align}
With this the master equation yields (with change of variables $t^{\prime} = t -\tau$)
\begin{align} 
  \frac{\partial}{\partial t} \hat{\rho}  (t)
 =&  g_{c}^2  \int\limits_{0}^{t} dt^{\prime} \;
         \left\{
                  \hat{L}  \bullet
                  \hat{L}^{\dag}
                     -
                  \hat{L}^{\dag}
                  \hat{L}  \bullet
          \right\}
          e^{ -   \frac{\kappa_c }{2}  t^{\prime} }
          \hat{\rho}(t- t^{\prime})
          + h.c.
\end{align} 
Under a Markov approximation we can evaluate the integral and for $t\rightarrow \infty$ we obtain:
\begin{align}\label{Eq.masterequationFinal}
  \frac{\partial}{\partial t} \hat{\rho}   
 =&    \kappa_{a} \hat{\mathcal{D}}
 \left[ \hat{L}   \right] \hat{\rho} ,
 \hspace{0.3cm}
\kappa_{a} =  \frac{4g_{c}^2}{\kappa_{c}}
\end{align}
here we use the superoperator definition
$\hat{\mathcal{D}}\left[ \hat{c} \right] \bullet = \hat{c} \bullet \hat{c}^{\dag} - \frac{1}{2} \hat{c}^{\dag} \hat{c} \bullet - \frac{1}{2} \bullet \hat{c}^{\dag} \hat{c}  $.
The master equation Eq.~\ref{Eq.masterequationFinal} contains the desired dissipative process.

\section{Effective nonlinear potential}\label{app_03}
\begin{figure}
    \centering
    \includegraphics[trim=0 0 23.5cm 0, clip, width=\linewidth]{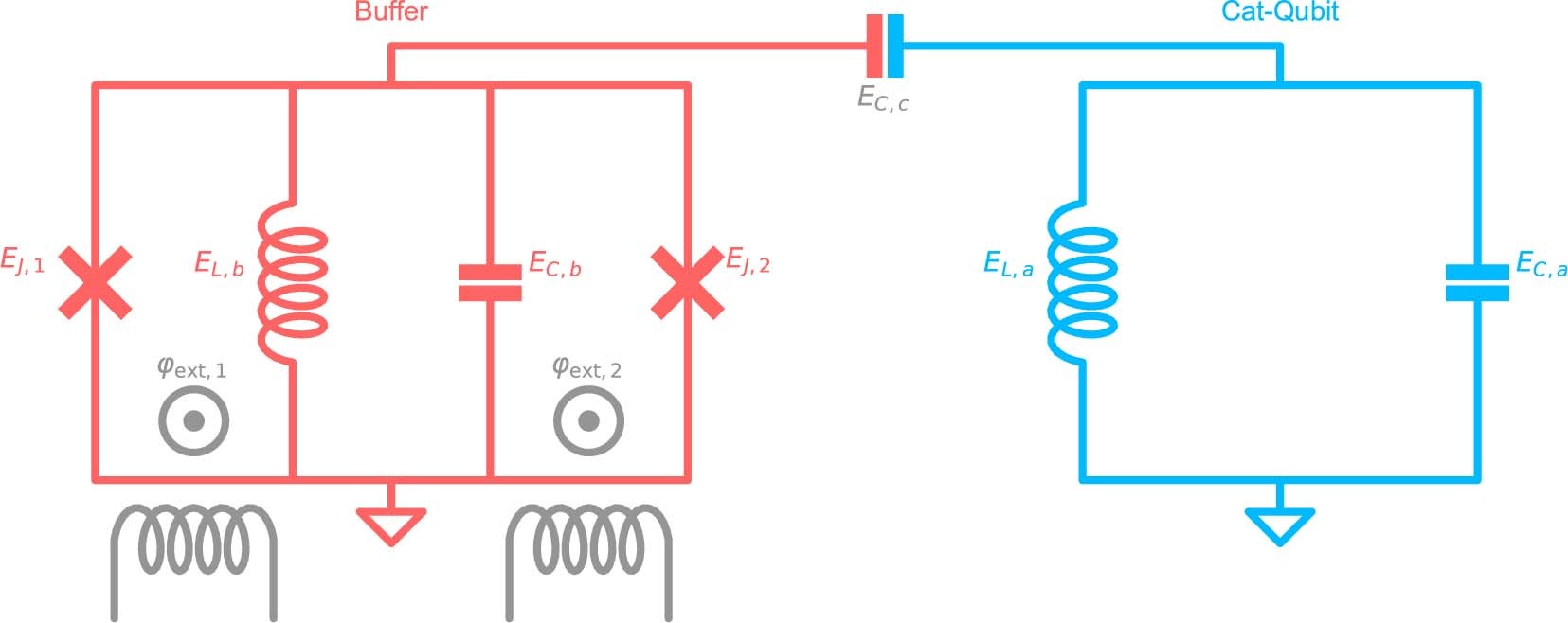}
    \caption{Sketch of ATS taken from \cite{LescanneNP20}. The ATS consists of two Josephson junctions with energies $E_{J,1}$ and $E_{J,2}$, shunted by an inductor with energy $E_{J,b}$ and a capacitor with energy $E_{C,b}$. There are two independent fluxes $\phi_{\mathrm{ext},1}$ and $\phi_{\mathrm{ext},2}$ passing through the left and right loops. The ATS can be capacitively coupled to the other four resonators required in our case, similar to the capacitive connection between red and blue parts.}
    \label{fig_ATS_sketch}
\end{figure}

In this appendix, we show how the nonlinear interactions in Eq.~\eqref{Eq03} can be implemented using superconducting circuit architectures with engineered flux nonlinearities. As a concrete platform, we consider an asymmetrically threaded SQUID (ATS) shown in Fig.~\ref{fig_ATS_sketch}, which provides a highly tunable nonlinear element.
The key advantage of the ATS is that external flux drives allow selective activation of higher-order nonlinear processes while suppressing undesired Kerr-type interactions, which we will illustrate later. The Hamiltonian of several superconducting resonators capacitively coupled to such an ATS device reads (alternatively, see Eq.~S2 in the supplementary materials of \cite{LescanneNP20})
\begin{align}
    \nonumber \hat{H}&=\hat{H}_0+\frac{1}{2}E_{L,b}\hat{\phi}^2-2E_J\cos(\phi_\Sigma)\cos(\hat{\phi}+\phi_\Delta)\\
    &\qquad +2\Delta E_J\sin(\phi_\Sigma)\sin(\hat{\phi}+\phi_\Delta), \label{Eq_ATS_001}
\end{align}
where $\hat{H}_0$ and $\hat{\phi}$ are the free Hamiltonian and flux operators that will be explicitly defined later in Eq.~\eqref{Eq_ATS_H_0} and \eqref{Eq_ATS_phi_operator}, $E_{L,b}$ is the inductive energy of the ATS, $E_J=(E_{J,1}+E_{J,2})/2$ is the average energy of the two Josephson junctions in ATS, and $\Delta E_J=(E_{J,1}-E_{J,2})/2$ is the energy deviation of the Josephson junctions from the average energy. Similarly, the fluxes $\phi_\Sigma=(\phi_{\mathrm{ext},1}+\phi_{\mathrm{ext},2})/2$ and $\phi_\Delta=(\phi_{\mathrm{ext},1}-\phi_{\mathrm{ext},2})/2$ denote the average and deviation from the average of the two external fluxes in Fig.~\ref{fig_ATS_sketch}. 

Since in our case, four superconducting resonators with their corresponding modes labeled as $a,b,c,d$ will be capacitively coupled to the ATS device, the free Hamiltonian $\hat{H}_0$ yields ($\hbar =1$)
\begin{align}
    \hat{H}_0&=\omega_a\hat{a}^\dag\hat{a}+\omega_b\hat{b}^\dag\hat{b}+\omega_c\hat{c}^\dag\hat{c}+\omega_d\hat{d}^\dag\hat{d}+\omega_e\hat{e}^\dag\hat{e},\label{Eq_ATS_H_0}
\end{align}
where the extra harmonic oscillator mode $\omega_e\hat{e}^\dag\hat{e}$ denotes the energy of the ATS mode. However, we note that this term can be ignored for the rest of our considerations by assuming it always stays in its ground state $\langle\hat{e}^\dag\hat{e}\rangle=0$, and that its frequency $\omega_e$ is sufficiently detuned from all other pump frequencies. That is, there will be no resonant nonlinear terms about the $e$-mode after going to a rotating frame and applying the rotating wave approximation (see \cite{HillmannPRA23}). Going to a rotating frame about the free Hamiltonian $\hat{H}_0$, Eq.~\eqref{Eq_ATS_001} becomes \begin{align}
    \nonumber \hat{H}&=\frac{1}{2}E_{L,b}\hat{\phi}^2-2E_J\cos(\phi_\Sigma)\cos(\hat{\phi}+\phi_\Delta)\\
    &\qquad +2\Delta E_J\sin(\phi_\Sigma)\sin(\hat{\phi}+\phi_\Delta), \label{Eq_ATS_01}
\end{align}
where the flux operator is defined as \begin{align}
    \hat{\phi}&=\sum_{\eta\in\{a,b,c,d\}}(\phi_\eta\hat{\eta}e^{-i\omega_\eta t}+h.c.),\label{Eq_ATS_phi_operator}
\end{align}
with $\phi_{\bullet}$ denoting the zero point fluctuation of each mode.

Moreover, to eliminate undesired Kerr-type nonlinearities arising from even powers of $\hat{\phi}$, we must engineer the driving configurations such that the Kerr-type nonlinear terms always appear with some explicit oscillating prefactors. This ensures that self- and cross-Kerr contributions acquire rapidly oscillating phases, and thus can be removed under the rotating wave approximation (RWA). Accordingly, we tune the external fluxes $\phi_{\mathrm{ext},1}$ and $\phi_{\mathrm{ext},2}$, such that $\phi_\Delta=0$ (or $\phi_{\mathrm{ext},1}=\phi_{\mathrm{ext},2}$), and
\begin{align}
    \phi_\Sigma&=\phi_{\mathrm{ext},\{1,2\}}=\frac{\pi}{2}+\sum_{k=1}^3(\epsilon_k e^{-i\omega_k t}+\epsilon_k^* e^{i\omega_kt})=\frac{\pi}{2}+\phi_\mathrm{dr},
\end{align}
meaning that both flux loops of the ATS are DC-biased  to $\pi/2$ and a multitone RF flux bias is applied.
Eq.~\eqref{Eq_ATS_01} then becomes \begin{align}
    \nonumber \hat{H}&=\frac{1}{2}E_{L,b}\hat{\phi}^2+2E_J\sin(\phi_\mathrm{dr})\cos(\hat{\phi})\\
    &\qquad +2\Delta E_J\cos(\phi_\mathrm{dr})\sin(\hat{\phi}).\label{Eq_ATS_02}
\end{align}
Here, we see that the even powers arising from the $\cos(\hat{\phi})$ term always acquire some driving frequencies from the odd powers from the $\sin(\phi_\mathrm{dr})$ term. This effectively avoids the unwanted mergence of Kerr-type interactions from the even term $\cos(\hat{\phi})$.

Expanding Eq.~\eqref{Eq_ATS_02} upto the second order in $\phi_\mathrm{dr}$, and fourth order in $\hat{\phi}$, we obtain \begin{align}
    \hat{H}&= \sum_{n=0}^{4} \; g_{n} \; \hat \phi^{n}, \label{Eq_ATS_08}
\end{align}
where \begin{align}
    \nonumber g_0&=2E_J\phi_\mathrm{dr},\\
    \nonumber g_1&=2\Delta E_J\left(1-\frac{1}{2}\phi_\mathrm{dr}^2\right),\\
    \nonumber g_2&=\frac{1}{2}E_{L,b}-E_J\phi_\mathrm{dr},\\
    \nonumber g_3&=\frac{1}{3}\Delta E_J\left(\frac{1}{2}\phi_\mathrm{dr}^2-1\right),\\
    g_4&=\frac{1}{12}E_J\phi_\mathrm{dr}. \label{Eq_ATS_04}
\end{align}
The resulting Hamiltonian thus contains wave-mixings processes up to the fourth-order. The constant term $g_0$ is dynamically irrelevant, for it is merely a shift of reference energy, and thus can be discarded. The linear term $g_1 \hat{\phi}$ would lead to an undesired displacement of the modes and hence, we will avoid resonant driving and can eliminated this contribution via a RWA later. The quadratic $g_2\hat{\phi}^2$ and quartic $g_4\hat{\phi}^4$ contributions are used to engineer $\hat{H}_1$, while the cubic term $g_3\hat{\phi}^3$ enables the realization of $\hat{H}_2$. Importantly, the latter requires $\Delta E_J \neq 0$, reflecting the necessity of junction asymmetry.

We now design the required frequency-matching conditions. For the first target interaction $\hat{H}_1$ of Eq.~\eqref{Eq03}, we have 
\begin{align}
    \hat{H}_1\sim\hat{a}^\dag(\hat{a}^\dag+\hat{b}^\dag\hat{b}-N)\hat{c}^\dag+h.c..
\end{align}
Since the middle term in brackets is Hermitian, it always acquires an overall zero oscillating frequency regardless of which rotating frame we chose. Hence, we only need to make sure that the one driving frequency $\omega_1$ matches the combined frequencies of the remaining $\hat{a}^\dag$ and $\hat{c}^\dag$ operators in the rotated frame with $\hat{H}_0$, i.e, we choose 
\begin{align}
    \omega_1&=\omega_a+\omega_c. \label{Eq_ATS_06}
\end{align}
This matching condition will induce the resonant terms $\sim \hat{a}^\dag\hat{c}^\dag$ from both the second $\hat{\phi}^2$ and fourth $\hat{\phi}^4$ order terms in Eq.~\eqref{Eq_ATS_04}, as well as $\sim \hat{a}^\dag\hat{a}^\dag\hat{a}\hat{c}^\dag$ and $\hat{a}^\dag\hat{b}^\dag\hat{b}\hat{c}^\dag$ from the fourth $\hat{\phi}^4$ order term.

To design the matching conditions for $\hat{H}_2$ in Eq.~\eqref{Eq03}, defined by \begin{align}
    \hat{H}_2\sim\hat{a}^2\hat{d}^\dag \pm \hat{b}^2\hat{d}^\dag+h.c., \label{Eq_ATS_05}
\end{align}
we similarly observe that in the rotated frame, Eq.~\eqref{Eq_ATS_05} catches two different oscillating frequencies of $\omega_d-2\omega_a$ and $\omega_d-2\omega_b$, which arise from the first and second terms respectively. At first glance, it might be tempting for us to use only one additional driving pump with the frequency $\omega_2$ such that $\omega_a=\omega_b$ and $\omega_2=\omega_d-2\omega_2=\omega_d-2\omega_b$. However, this choice is actually problematic, for the condition $\omega_a=\omega_b$ would symmetrically induce another resonant interaction term in parallel to $\hat{H}_1$ \begin{align}
    \hat{H}_{1,b}&\sim \hat{b}^\dag(\hat{a}^\dag\hat{a}+\hat{b}^\dag\hat{b}-N)\hat{c}^\dag+h.c..
\end{align}
Consequently, we need to think about using two driving pumps $\omega_2$ and $\omega_3$ to separately match the two frequencies induced by Eq.~\eqref{Eq_ATS_05}, i.e., \begin{align}
    2\omega_2&=\omega_d-2\omega_a, \qquad 2\omega_3=\omega_d-2\omega_b. \label{Eq_ATS_07}
\end{align}
Those driving frequencies ensure the freedom of choosing the $\pm$ sign in Eq.\eqref{Eq_ATS_05}. However, such a driving configuration would produce extra unwanted resonance terms, e.g., $\sim\epsilon_2\epsilon_3\hat{a}\hat{b}\hat{d}^\dag$. Nevertheless, the prefactor $g_3$ for the cubic term $\hat{\phi}^3$ given in Eq.~\eqref{Eq_ATS_04} is quadratic in the driving field 
$\phi_\mathrm{dr}$ and thus $g_3$ is proportional to 
\begin{align}
    \nonumber \phi_\mathrm{dr}^2&=(\epsilon_1e^{-i\omega_1t}+\epsilon_2 e^{-i\omega_2t}+\epsilon_3e^{-i\omega_3t}+h.c.)^2,\\
    &=\sum_{m,n=1}^3 (\epsilon_m\epsilon_n e^{-i(\omega_m+\omega_n)t}+\epsilon_m\epsilon_n^* e^{-i(\omega_m-\omega_n)t}+h.c.),
\end{align}
which consists of all possible combinations of driving frequencies $\omega_m\pm\omega_n$, not only $2\omega_2=\omega_2+\omega_2$, and we can use any other combinations to realize the parametric processes for $\hat{H}_2$.
With this, we find that by choosing 
\begin{align}
    \nonumber \omega_2+\omega_1&=\omega_d-2\omega_a,\\
    \omega_3+\omega_1&=\omega_d-2\omega_b,
\end{align}
the induced resonant terms are precisely only the required ones. We note that alternative driving schemes in such a multi-wave mixing architecture might also be possible but not explored further here.

As a short summary, we have proposed three frequency matching conditions using three driving pumps \begin{align}
    \nonumber \omega_1 &= \omega_a + \omega_c,\\
    \nonumber \omega_2 + \omega_1 &= \omega_d - 2\omega_a,\\
    \omega_3 + \omega_1 &= \omega_d - 2\omega_b,
\end{align}
where the first line ensures that the necessary terms needed for $\hat{H}_1$ are on resonance, the second and third lines ensures that of $\hat{H}_2$ on resonance. Specifically, the surviving terms induced by these frequency matching conditions after applying RWA can be divided into three parts \begin{align}
    \hat{H}=\hat{H}_1'+\hat{H}_2'+\hat{H}_3', \label{Eq_ATS_H_rwa}
\end{align}
where $\hat{H}_1'$ and $\hat{H}_2'$ are the responsible parts for realizing $\hat{H}_1$ and $\hat{H}_2$ in Eq. \eqref{Eq03}, and $\hat{H}_3'$ is an extra term that has no effective contributions (i.e., $\hat{H}_3'\equiv 0$) in the limit of strong single photon loss rates $\kappa_c$ and $\kappa_d$.

We first analyze $\hat{H}_1'$, which can be rewritten into a simpler form as 
\begin{align}
    \nonumber \hat{H}_1'&=
    A \; 
    \hat{a}^\dag
    (\hat{a}^\dag\hat{a}+B\hat{b}^\dag\hat{b}-C)\hat{c}^\dag+h.c.,
\end{align}
whit the coefficients
\begin{align}
    \nonumber A&= E_J \phi_c \phi_a^3\epsilon_1,
    \hspace{1cm}
    \nonumber B =2\phi_b^2 / \phi_a^2 ,
    \\
    C&= 2/\phi_a^2-(\phi_a^2+\phi_b^2+\phi_c^2+\phi_d^2)/\phi_a^2   .
\end{align}
If we further fine tune the system parameters $\phi_\bullet$ and $\epsilon_\bullet$, such as $\phi_a=\sqrt{2}\phi_b$ to ensure $B = 1$, as well as $C=N$, we will obtain the desired nonlinear interaction 
\begin{align}
    \hat{H}_1&= A \; 
    \hat{a}^\dag(\hat{a}^\dag\hat{a}+\hat{b}^\dag\hat{b}-N)\hat{c}^\dag+h.c.. \label{Eq05}
\end{align}
Using the parameters reported in \cite{LescanneNP20, RegladeN24, MarquetPRX24} as a reference, we note that the engineered asymmetries in the zero-point fluctuations $\phi_{a,b}$ and mode frequencies $\omega_{a,b}$ of coupled resonators are experimentally feasible. In particular, the condition $\phi_a = \sqrt{2}\,\phi_b$ can be satisfied within realistic circuit-QED parameter regimes through appropriate impedance engineering of the respective modes. Furthermore, in the regime where $\phi_\bullet$ are comparably small $\sim 0.01-0.1$ as shown in Table~S1 of \cite{RegladeN24}, the term $-(\phi_a^2+\phi_b^2+\phi_c^2+\phi_d^2)/\phi_a^2$ in $C$ would contribute approximately a factor of $-4$. This can be compensated by tuning $\phi_a$ sufficiently small, such that the first term $2/ \phi_a^2\approx N+4\gg 1$ becomes an integer. The fine tuning of $\phi_\bullet$ has been shown to be possible upto the precision of $10^{-2}$ \cite{RegladeN24} to $10^{-4}$ \cite{MarquetPRX24}. Although using a small $\phi_a$ to engineer a large integer $N$ would simultaneously weaken the coupling strength, i.e., $A\sim\phi_a^3\epsilon_1$, we may still use a sufficently large driving amplitude $\epsilon_1$ to maintain a strong coupling.

Similarly, to realize $\hat{H}_2$, we need to analyze $\hat{H}_2'$ that can be written as 
\begin{align}
    \hat{H}_2'&=(A'\hat{a}^2\pm B'\hat{b}^2)\hat{d}^\dag+h.c.,
\end{align}
where \begin{align}
    A'=2\Delta E_J \phi_b^2\phi_d\epsilon_1\epsilon_2, \qquad B'=\Delta E_J \phi_b^2\phi_d\epsilon_1\epsilon_3.
\end{align}
here we already accounted for the condition $\phi_{a} = \sqrt{2} \phi_{b}$. By tuning the driving amplitudes to $2\epsilon_2=\pm\epsilon_3$, we obtain the desired Hamiltonian \begin{align}
    \hat{H}_2&=A'(\hat{a}^2\pm\hat{b}^2)\hat{d}^\dag+h.c..
\end{align}

To eventually ensure that the unwanted residual terms $\hat{H}_3'$ does not influence the circuit dynamics, we first note that it consists of the following terms \begin{align}
    \hat{H}_3'&=E_{L,b}\sum_{\eta\in\{a,b,c,d\}}\phi_\eta^2\left(\hat{\eta}^\dag\hat{\eta}+\frac{1}{2}\right)+\hat{G}\hat{d}^\dag\hat{d}
\end{align}
where $\hat{G}=2E_J\phi_a\phi_c\phi_d^2(\epsilon_1\hat{a}^\dag\hat{c}^\dag+\epsilon_1^*\hat{a}\hat{c})$. We observe that the first term is essentially a shift in the resonance frequencies of the resonators, which can be easily eliminated by initially choosing a slightly different rotating frame with the detuned frequencies $\omega_\bullet'$, e.g., $\omega_a'=\omega_a-E_{L,b}\phi_a^2$. The second term effectively behaves as the zero operator $\hat{0}$ in the limit of sufficiently large $\kappa_d$, for in this limit, the resonator $d$ mimics a Markovian bath that remains always in the vacuum state $|0\rangle_d$. Consequently, the second term $\sim\hat{d}^\dag\hat{d}$ functions as the zero operator, i.e., $\hat{d}^\dag\hat{d}|0\rangle_d\otimes|\psi\rangle_{a,b,c}=0$ regardless of states in other modes, and therefore does not affect the system dynamics constrained in the $|0\rangle_d$ manifold.

\section{Fidelity as a function of noise strength}\label{app_04}
In this section, we investigate how the steady state fidelity of the system with respect to the target state behaves for different noise channels. In particular, we show that the fidelity in the presence of the single-photon loss channel in the $b$-mode always decays towards around $\frac{1}{2}$ regardless of large or small loss rate $\kappa_b$. In fact, the loss rate $\kappa_b$ only determines how fast the fidelity decays towards $\frac{1}{2}$.
\begin{figure}
    \centering
    \includegraphics[width=\linewidth]{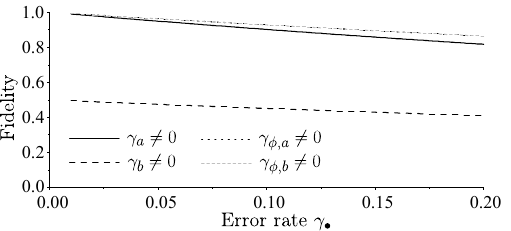}
    \caption{Fidelity of the dissipatively prepared $\mathfrak{su}(2)$ cat state $|\psi_+\rangle$ in the presence of different types of noise: $\sqrt{\gamma_a}\hat{a}$, $\sqrt{\gamma_b}\hat{b}$, $\sqrt{\gamma_{\phi,a}}\hat{a}^\dag\hat{a}$, and $\sqrt{\gamma_{\phi,b}}\hat{b}^\dag\hat{b}$. We set $\hat{H}=0$, $N=8$, and $\kappa_1=\kappa_2=1$ for simplicity. The fidelity approaches to $\frac{1}{2}$ for the case of $\gamma_b\neq 0$ because $\hat{b}$ can induce the phase-flip error within the steady state manifold, which cannot be corrected by the engineered dissipation.}
    \label{fig_fid_gamma_dependence}
\end{figure}
In the presence of an unwanted noise source $\sqrt{\gamma'}\hat{L}'$, we may denote its Lindblad equation as \begin{align}
    \nonumber \frac{d}{dt}\hat{\rho}(t)&=\big[\underbrace{\kappa_1\hat{\mathcal{D}}(\hat{L}_1)+\kappa_2\hat{\mathcal{D}}(\hat{L}_2)}_{\hat{\mathcal{L}}}+\underbrace{\gamma'\hat{\mathcal{D}}(\hat{L}')}_{\hat{\mathcal{L}}'}\big]\hat{\rho}(t),\\
    &=(\hat{\mathcal{L}}+\hat{\mathcal{L}}')\hat{\rho}(t), \label{eq_general_noisy_model}
\end{align}
where $\gamma'\ll \kappa_1,\kappa_2$.

Going to an interaction picture about $\hat{\mathcal{L}}$ and then back to the Schr\"odinger picture, we can write the solution for Eq.~\eqref{eq_general_noisy_model} as \begin{align}
    \hat{\rho}(t)=\hat{\mathcal{S}}(t)\hat{\rho}(0),
\end{align}
where $\hat{\mathcal{S}}$ is the superoperator describing the open system evolution of the system initialized in $\hat{\rho}(0)$, and is written as \begin{align}
    \hat{\mathcal{S}}(t)&=e^{\hat{\mathcal{L}}t}\overleftarrow{\mathcal{T}}e^{\int_0^t e^{-\hat{\mathcal{L}}t'}\hat{\mathcal{L}}'e^{\hat{\mathcal{L}}t'}dt'}, \label{eq_evolution_interaction_pic}
\end{align}
which is Eq.~18 in \cite{IppolitiPRA15}.

For simplicity, we set $\kappa_1=\kappa_2=1$. We expand the exponential term in  Eq.~\eqref{eq_evolution_interaction_pic} and then insert the identity operator $\mathds{1}=\hat{\mathcal{R}}+\hat{\mathcal{Q}}$ in between $e^{-\hat{\mathcal{L}}t'}$ and $\hat{\mathcal{L}}'$. Here, the projector $\hat{\mathcal{R}}=\lim_{t\rightarrow\infty}e^{\hat{\mathcal{L}}t}$ projects any state to the steady state manifold $\mathcal{H}_0$ of the lossless dynamics Eq.~\eqref{eq_general_noisy_model}, which in our case is spanned by the two-mode binomial coherent states $|N,\pm1\rangle$, and $\hat{\mathcal{Q}}=1-\hat{\mathcal{R}}$ projects to the orthogonal complement $\mathcal{H}_0^\perp$ of the steady state manifold. We see that these operators indeed obey the following projector identities \begin{align}
    \hat{\mathcal{R}}^2=\hat{\mathcal{R}}, \quad \hat{\mathcal{Q}}^2=\hat{\mathcal{Q}}, \quad \hat{\mathcal{R}}\hat{\mathcal{Q}}=\hat{\mathcal{Q}}\hat{\mathcal{R}}=0.
\end{align}
The detailed discussion about the projector formalism using the above defined projectors can be found in \cite{PopkovPRA18}. The expansion result is \begin{align}
    \nonumber \hat{\mathcal{S}}(t)&=e^{\hat{\mathcal{L}}t}+e^{\hat{\mathcal{L}}t}\int_0^t e^{-\hat{\mathcal{L}}t'}\hat{\mathcal{L}}'e^{\hat{\mathcal{L}}t'}dt'\\
    \nonumber &\qquad+e^{\hat{\mathcal{L}}t}\int_0^t e^{-\hat{\mathcal{L}}t_1}\hat{\mathcal{L}}'e^{\hat{\mathcal{L}}t_1}dt_1\int_0^{t_1} e^{-\hat{\mathcal{L}}t_2}\hat{\mathcal{L}}'e^{\hat{\mathcal{L}}t_2}dt_2+...\\
    \nonumber &=e^{\hat{\mathcal{L}}t}+e^{\hat{\mathcal{L}}t}\int_0^t e^{-\hat{\mathcal{L}}t_1}(\hat{\mathcal{R}}+\hat{\mathcal{Q}})\hat{\mathcal{L}}'e^{\hat{\mathcal{L}}t_1}dt_1(\hat{\mathcal{R}}+\hat{\mathcal{Q}})\\
    \nonumber &\qquad +e^{\hat{\mathcal{L}}t}\int_0^t e^{-\hat{\mathcal{L}}t_1}(\hat{\mathcal{R}}+\hat{\mathcal{Q}})\hat{\mathcal{L}}'e^{\hat{\mathcal{L}}t_1}dt_1\\
    \nonumber &\qquad \times\int_0^{t_1} e^{-\hat{\mathcal{L}}t_2}(\hat{\mathcal{R}}+\hat{\mathcal{Q}})\hat{\mathcal{L}}'e^{\hat{\mathcal{L}}t_2}dt_2(\hat{\mathcal{R}}+\hat{\mathcal{Q}})+...\\
    &=e^{\hat{\mathcal{L}}t}+\sum_{n=1}^\infty t^n\frac{(\hat{\mathcal{R}}\hat{\mathcal{L}}'\hat{\mathcal{R}})^n}{n!}+\mathcal{O}(\hat{\mathcal{Q}}), \label{eq_dyson_series_expansion}
\end{align}
where we have only grouped the confined dynamical terms containing only $\hat{\mathcal{R}}$, and $\mathcal{O}(\hat{\mathcal{Q}})$ denote the leakage terms to the subspace $\hat{\mathcal{Q}}$. More discussions about why these contributions from leakage terms are small enough to be neglected can be found in \cite{ZanardiPRL14}.

The fidelity between the system initialized in $\hat{\rho}(0)=\hat{\rho}_0=|\psi_0\rangle\langle \psi_0|$ and the target state $|\psi_{+,N}\rangle\sim|N,1\rangle+|N,-1\rangle$, where $|\psi_0\rangle$ is the initial state denoted in (7) of Table.~\ref{table_summary_jump_operators} for $N-m=\text{even}$, and $|\psi_{+,N}\rangle$ is the even two-mode binomial cat state, evolves as \begin{align}
    \nonumber \mathcal{F}(t)&=\langle\psi_{+,N}|\hat{\mathcal{S}}(t)\hat{\rho}_0|\psi_{+,N}\rangle,\\
    &\approx \langle \psi_{+,N}|e^{\hat{\mathcal{L}}t}\hat{\rho}_0|\psi_{+,N}\rangle + \sum_{n=1}^\infty t^n\frac{\langle \psi_{+,N}|(\hat{\mathcal{R}}\hat{\mathcal{L}}'\hat{\mathcal{R}})^n\hat{\rho}_0|\psi_{+,N}\rangle}{n!}.\label{eq_approx_fid_evolution}
\end{align}
To evaluate Eq.~\eqref{eq_approx_fid_evolution} for $\hat{\mathcal{L}}'$ being the single-photon loss in the $b$-mode, we first define $\hat{\rho}_{\pm,N}=|\psi_{\pm,N}\rangle\langle \psi_{\pm,N}|$, and then note the following useful formulas \begin{align}
    \nonumber \hat{b}^\dag|\psi_{\pm,N}\rangle&\approx\sqrt{\frac{N}{2}}|\psi_{\pm,N+1}\rangle,\\
    \nonumber \hat{\mathcal{D}}(\sqrt{\gamma'}\hat{b})\hat{\rho}_{\pm,N}&\approx \frac{\gamma'N}{2}(\hat{\rho}_{\pm,N-1}-\hat{\rho}_{\pm,N}),\\
    \hat{\mathcal{R}}\hat{\rho}_{\pm,N-1}&=\hat{\rho}_{\mp,N},
\end{align}
where the first line can be understood analogously to that of the single-mode coherent state $\hat{a}^\dag|\alpha\rangle\approx \alpha^*|\alpha\rangle$ for sufficiently large $|\alpha|\equiv\sqrt{N/2}$. With this, we can evaluate \begin{align}
    (\hat{\mathcal{R}}\hat{\mathcal{L}}'\hat{\mathcal{R}})^n\hat{\rho}_0&=\frac{1}{2}(-\gamma'N)^n(\hat{\rho}_{\pm,N}-\hat{\rho}_{\pm,N}).
\end{align}
Thus, Eq.~\eqref{eq_approx_fid_evolution} becomes \begin{align}
    \nonumber \mathcal{F}(t)&\approx \frac{1}{2}+\frac{1}{2}e^{-\gamma'Nt}-e^{-\frac{N}{2}t}(1-|\langle\psi_0|\psi_{+,N}\rangle|^2),\\
    &\approx \frac{1}{2}+\frac{1}{2}e^{-\gamma'Nt}-e^{-\frac{N}{2} t}. \label{eq_fid_approx_b}
\end{align}
Here, we have also approximated the lossless fidelity evolution term as a simple recovery map applied stochastically with a characteristic time $\frac{2}{N}$ (see Eq.~05 in \cite{IppolitiPRA15}) \begin{align}
    e^{\hat{\mathcal{L}}t}\approx e^{-\frac{N}{2}t}\mathds{1}+(1-e^{-\frac{N}{2}t})\hat{\mathcal{R}}. \label{eq_stochastic_approx}
\end{align}
This is valid because we are working in the region $\gamma'\ll\kappa_1,\kappa_2$, where the leakage to the $\hat{\mathcal{Q}}$ subspace is very small. Here, we assume the characteristic time being proportional to $\frac{1}{N\kappa}$ because the jump operator $\hat{L}_2$ has a truncated operator norm $\norm{\hat{L}_2}_{\mathcal{H}_0}=N$, which induces an effective decay rate of $N\kappa$. The usage of truncated operator norm is a common practice in approximating the ``strength" of an operator, see \cite{Winter17, ShirokovMs20, Becker25}. We observe that Eq.~\ref{eq_fid_approx_b} converges to $\frac{1}{2}$ in the long time limit, indicating that in the presence of the single-photon loss $\hat{b}$, the fidelity would exponentially converge to $\frac{1}{2}$ at an effective rate of $N\gamma'$.

Whereas if $\hat{\mathcal{L}}'$ is the single-photon loss in the $a$-mode, or dephasings in both $a$ and $b$ modes, we will instead have $\hat{\mathcal{R}}\hat{\mathcal{L}}'\hat{\mathcal{R}}\approx 0$ for sufficiently large $N$. This can be seen, using the single-photon loss in the $a$-mode as an example, \begin{align}
    \nonumber \hat{\mathcal{R}}\hat{\mathcal{D}}(\sqrt{\gamma'}\hat{a})\hat{\mathcal{R}}\hat{\rho}_0&=\hat{\mathcal{R}}\hat{\mathcal{D}}(\sqrt{\gamma'}\hat{a})\hat{\rho}_{+,N},\\
    \nonumber &\approx\frac{\gamma'N}{2}\hat{\mathcal{R}}(\hat{\rho}_{\mp,N-1}-\hat{\rho}_{\pm,N}),\\
    \nonumber &=\frac{\gamma'N}{2}(\hat{\rho}_{\pm,N}-\hat{\rho}_{\pm,N}),\\
    &=0.
\end{align}
Therefore, Eq.~\eqref{eq_approx_fid_evolution} becomes simply \begin{align}
    \mathcal{F}(t)&\approx 1-e^{-\frac{N}{2} t}(1-|\langle\psi_0|+\rangle|^2).\label{eq_fid_approx_c}
\end{align}
Taking the steady state limit $t\rightarrow\infty$, Eq.~\eqref{eq_fid_approx_b} and \eqref{eq_fid_approx_c} converge to $\frac{1}{2}$ and $1$ respectively, which represents the steady state fidelity with, and without, the single-photon loss in $b$-mode. This can be seen from Fig.~\ref{fig_fid_gamma_dependence}, where the steady state fidelities with and without $\gamma_b$ are approximately horizontal lines with fidelities $1$ and $\frac{1}{2}$ with negligible deviations. However, as the unwanted noise strength becomes increasingly large, that the limit $\gamma'\ll\kappa_1,\kappa_2$ no longer holds, the leakage from the target state manifold $\hat{\mathcal{R}}$ into the outside manifold $\hat{\mathcal{Q}}$ increases. Hence, the steady state fidelity would deviate more from their ideal values as shown in Fig.~\ref{fig_fid_gamma_dependence} for larger $\gamma_\bullet$.

\end{document}